\documentclass[12pt]{iopart}
\usepackage{graphicx}
\usepackage{amssymb}

\begin{document}

\title[Tl2201 Brings Spectroscopic Probes Deep into the Overdoped
Regime]{Tl$_2$Ba$_2$CuO$_{6+\delta}$ Brings Spectroscopic Probes Deep Into the Overdoped Regime of the High-T$_c$ Cuprates}

\author{DC Peets, JDF Mottershead, B Wu, IS Elfimov,
R Liang, WN Hardy, DA Bonn, M Raudsepp$^{\dag}$,  NJC Ingle and A Damascelli}
\address{Department of Physics and Astronomy, University of
British Columbia, \\ Vancouver, BC V6T 1Z1 Canada}
\address{\dag\ Department of Earth and Ocean Sciences, University of
British Columbia, Vancouver, BC V6T 1Z4 Canada} \ead{bonn@physics.ubc.ca and damascelli@physics.ubc.ca}

\begin{abstract}
Single-particle spectroscopic probes, such as scanning tunneling and angle-resolved photoemission spectroscopy (ARPES), have provided us with
crucial insights into the complex electronic structure of the high-$T_c$ cuprates, in particular for the under and optimally doped regimes where
high-quality crystals suitable for surface-sensitive experiments are available. Conversely, the elementary excitations on the heavily overdoped
side of the phase diagram remain largely unexplored. Important breakthroughs could come from the study of Tl$_2$Ba$_2$CuO$_{6+\delta}$ (Tl2201),
a structurally simple system whose doping level can be tuned from optimal to extreme overdoping by varying the oxygen content. Using a self-flux
method and encapsulation, we have grown single crystals of Tl2201, which were then carefully annealed under controlled oxygen partial pressures.
Their high quality and homogeneity are demonstrated by narrow rocking curves and superconducting transition widths. For higher dopings, the
crystals are orthorhombic, a lattice distortion stabilized by O interstitials in the TlO layer. These crystals have enabled the first successful
ARPES study of both normal and superconducting-state electronic structure in Tl2201, allowing a direct comparison with the Fermi surface from
magnetoresistance and the gap from thermal conductivity experiments. This establishes Tl2201 as the first high-$T_c$ cuprate for which a
surface-sensitive single-particle spectroscopy and a comparable bulk transport technique have arrived at quantitative agreement on a major
feature such as the normal state Fermi surface. The momentum dependence of the ARPES lineshape reveals, however, an unexpected phenomenology: in
contrast to the case of under- and optimally-doped cuprates, quasiparticles are sharp near $(\pi,0)$, the antinodal region where the gap is
maximum, and broad at $(\pi/2,\pi/2)$, the nodal region where the gap vanishes. This reversed quasiparticle anisotropy past optimal doping, and
its relevance to scattering, many-body, and quantum-critical phenomena in the high-$T_c$ cuprates, are discussed.

\vspace{1.25cm} \noindent {\bf Submitted to New Journal of Physics (September 4, 2006).}

\end{abstract}

\maketitle

\section{Introduction}

\begin{figure}[t]
\begin{center}
\includegraphics[width=4.1in]{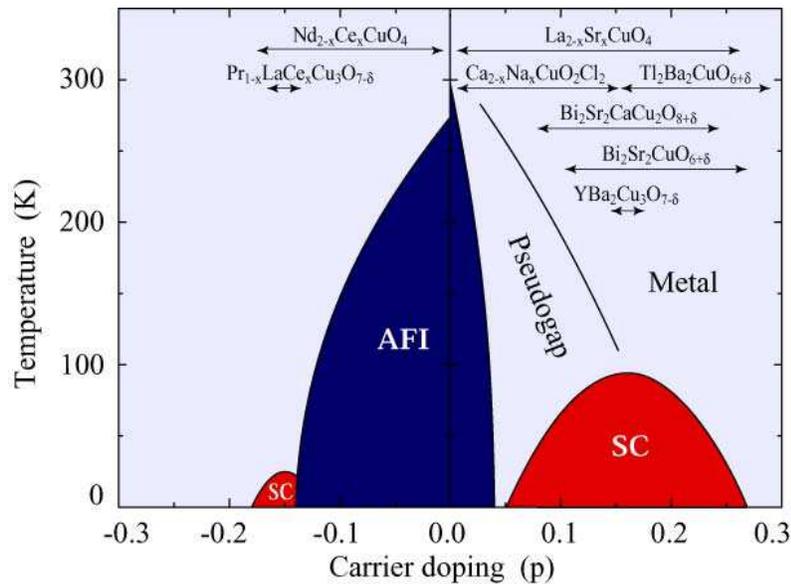}
\caption{\label{doping-diagram}Generic temperature-doping phase diagram for both electron ($p\!<\!0$) and hole-doped  ($p\!>\!0$) cuprates. The
doping range actually explored by ARPES for those material families more extensively studied is indicated by the corresponding arrows.}
\end{center}
\end{figure}

The cuprate superconductors can be tuned through a remarkable progression of states of matter by doping charge carriers into the CuO$_2$ planes
\cite{Bonn:2006}. The most generic feature of this tuning is a sequence from an antiferromagnetic insulator, through a doping range where the
superconducting critical temperature builds to a maximum (optimal doping) and then dies away again at higher doping, in the so-called overdoped
regime (Fig.\,\ref{doping-diagram}). This sequence contains three separate touchstones for an understanding of the cuprates:  the Mott
insulator, which is known to result from very strong electron-electron repulsion and is antiferromagnetically ordered, the $d$-wave
superconductor, and the overdoped metal. The latter is widely believed to exhibit less exotic normal-state properties and might be understood
through Fermi liquid theory. Over the last two decades, a great deal of experimental work has been done on undoped, underdoped, and optimally
doped cuprates, aiming at elucidating the connection between the antiferromagnetic insulator and high-$T_c$ superconductor (HTSC), and the role
of electronic correlations in the emergence of superconductivity. However, the testing of Fermi liquid theory in the overdoped regime has been
severely hampered by a lack of compounds that can actually be doped to this high level and are suitable for a wide range of experimental
techniques.

In the struggle to understand high-$T_c$ superconductivity, the quest for better materials to study remains key. Researchers want samples that
can be grown very cleanly, can be doped over a very wide range of the phase diagram, and are suitable for a variety of bulk and surface
measurements. Surface-sensitive techniques in particular, such as angle-resolved photoemission spectroscopy (ARPES) \cite{Andrea:2003,Campuzano}
and scanning tunneling microscopy (STM) and spectroscopy (STS) \cite{STM_RMP}, introduce additional constraints with regard to sample
requirements. These highly sophisticated spectroscopic probes can be used to obtain information on the single-particle excitation spectrum of a
solid --- i.e.\ the spectral function $A({\bf k},\omega)$ that can be calculated in terms of the electron Green's function starting from the
microscopic many-body Hamiltonian of the system \cite{AndreaScripta} --- but only if the material can be cleaved to expose a well-defined and
stable surface (i.e. which does not reconstruct, obscuring the bulk electronic structure).

A summary of the approximate doping range explored by ARPES on different HTSC families is given in Fig.\ref{doping-diagram}. On the undoped side
of the phase diagram ($p\!=\!0$), the parent compounds Ca$_2$CuO$_2$Cl$_2$ and Sr$_2$CuO$_2$Cl$_2$ yield excellent cleaved [001] faces that have
allowed ARPES studies of the electronic structure of the correlated antiferromagnetic insulator and of the magneto-elastic interactions
associated with the propagation of a single hole in the two-dimensional spin background \cite{KimC:Systps,ronning:1,Kyle:eph}. Via hole doping,
one can drive these materials through a metal-insulator transition and enter the superconducting dome (in this paper we will focus on the
hole-doped side of the phase diagram). For instance, recent STM, STS \cite{3}, and ARPES \cite{2} experiments on the underdoped superconductor
Ca$_{2-x}$Na$_x$CuO$_2$Cl$_2$ have suggested the existence of new electronic ordering phenomena in the doping region between the insulator and
the superconductor, possibly in the form of glassy electronic behavior or a surface-nucleated phase transition as more conventional long-range
charge order was not observed in scattering studies \cite{AbbamonteNCCOC}.
Near optimal doping, where the cuprates exhibit their highest superconducting critical temperatures, YBa$_2$Cu$_3$O$_{7-\delta}$ stands out as
the cleanest material and has been used extensively in bulk-sensitive studies of the normal and superconducting properties, such as the symmetry
of the order parameter \cite{5,TsueiCC:Paiscs,KirtleyNew}, superfluid density \cite{bonnandhardy}, charge and thermal transport \cite{4},
low-energy electrodynamics \cite{TimuskRMP,TomRMP}, vibrational and magnetic excitation spectra \cite{neutron1,neutron2}. Unfortunately, this
material is complicated by the presence of CuO-chain layers and does not have a neutral [001] cleavage plane for surface-sensitive techniques.
In particular, YBa$_2$Cu$_3$O$_{7-\delta}$ cleaves between the chain layer and the BaO layer. BaO surfaces have had the nearest source of
dopants (the CuO chains) removed, so that their hole-doping is uncertain. STS shows that the CuO chain surfaces are characterized by prominent
surface density waves \cite{8} and differ substantially from the bulk. They exhibit surface states, as seen in ARPES \cite{ schabel}, and
possess unavoidable doping disorder. These problems have severely limited the availability of single-particle spectroscopy data from ARPES
\cite{7,boriYBCO}, STM, and STS \cite{8,Renner}.

Extensive ARPES investigations have been performed on La$_{2-x}$Sr$_x$CuO$_{4}$ \cite{Zhoureview}, over a broad doping range extending from the
undoped insulator ($x\!=\!0$) to the overdoped superconductor ($x\!=\!0.22$; note that  $x\!=\!p$ for this compound). This system, however, is
affected by severe cation disorder associated with La-Sr substitution right next to the CuO$_2$ plane, as well as lattice, charge, and magnetic
instabilities, which might provide an explanation for why the maximum $T_c$ in this compound is suppressed with respect to other single-layer
materials \cite{Eisaki:2004}. In turn, many of the key bulk-sensitive measurements requiring perfect crystals and long mean free paths could not
be performed for this material family. In addition, neither STM nor STS experiments have been successful so far on cleaved
La$_{2-x}$Sr$_x$CuO$_{4}$, probably because both the CuO$_2$ and La$_{1-\frac{x}{2}}$Sr$_{\frac{x}{2}}$O surfaces are polar, with a charge
$z_{{\rm Cu}{\rm O}_2}\!=\!-(2\!-\!x)$ and $z_{{\rm (La,Sr)O}}\!=\!(1\!-\!x/2)$, and thus critically unstable. To date, the most rich and
comprehensive spectroscopic information on high-$T_c$ cuprates from both STS and ARPES have been obtained on the Bi compounds, and in particular
Bi$_2$Sr$_2$CuO$_{6+\delta}$ and Bi$_2$Sr$_2$CaCu$_2$O$_{8+\delta}$ \cite{Andrea:2003,Campuzano,STM_RMP}. However, similar to
La$_{2-x}$Sr$_x$CuO$_{4}$, this family has greater problems with disorder than YBa$_2$Cu$_3$O$_{7-\delta}$ \cite{Eisaki:2004}, leaving a
disconnection between materials for which we have the most information. This dilemma interferes with strict tests of the theory of high-$T_c$
superconductivity, in which one must connect the single particle excitations to the other physical properties and dynamic susceptibilities of
the cuprates \cite{11}.

A further problem in the study of HTSCs is that, due to the high $T_c$ values, the physical properties of the underlying normal state can be
probed only at relatively high temperatures. This often makes the detailed interpretation of the experimental results more complex and less
informative (e.g., charge and thermal conductivity, Hall number, magnetoresistance, etc.), or can even make some experiments completely
unfeasible (as in the case of the de Haas-van Alphen effect). Alternatively, at least for certain techniques, one can access the low-temperature
normal-state properties by suppressing superconductivity via the application of an external magnetic field. Once again however, due to the very
high $T_c$ and in turn the extraordinarily large value of the second critical field $H_{c2}$, this approach is precluded near optimal doping on
the cleanest cuprates. Extreme overdoping is a means of reducing both $T_c$ and $H_{c2}$, so that one can study both the superconducting and
underlying normal states, but here most samples tend to be less clean because of the need to dope with large concentrations of cation
impurities, such as Sr in La$_2$CuO$_{4}$ and Pb in Bi$_2$Sr$_2$CuO$_{6+\delta}$ \cite{Eisaki:2004}. In this case it can be unclear whether
changes observed in the material's properties are attributable to the increased doping, or the increased scattering and/or other extrinsic
effects. Due to these material and experimental limitations, the underlying normal state of HTSCs, the heavily overdoped side of the HTSC phase
diagram, and the metallic state that emerges beyond the end of the superconducting dome have remained so far largely unexplored
(Fig.\ref{doping-diagram}).

The most promising compound for investigating the deeply overdoped regime is the single-layer system Tl$_2$Ba$_2$CuO$_{6+\delta}$ (Tl2201, see
Fig.\ \ref{Tl-structure}), whose natural doping range varies from optimal to extreme overdoping as one increases the oxygen content $\delta$.
Tl2201 has two extremely flat CuO$_2$ planes located far apart from each other in the body-centered unit cell; there are no complicating chain
layers (as opposed to the Y-based cuprates), or bilayer band-splitting effects (as opposed to Bi$_2$Sr$_2$CaCu$_2$O$_{8+\delta}$ and
YBa$_2$Cu$_3$O$_{7-\delta}$). Like La and Bi-based systems, Tl2201 can exhibit a cation non-stoichiometry but this consists in excess Cu
substituting Tl in the Tl$_2$O$_2$ double layer, farther removed from the CuO$_2$ planes than the Bi-Sr or La-Sr substitutions that have been
shown to have a large impact on the superconductivity of the Bi and La cuprates \cite{Eisaki:2004}. Tl2201 may be reversibly doped through
interstitial oxygen and, near optimal doping, vacancies in the Tl$_2$O$_2$ bilayer \cite{Wagner}. The ability to overdope a crystal reversibly
from optimal doping to the full suppression of superconductivity, by changing its oxygen content alone, is a fortuitous property of Tl2201,
although this has been successfully demonstrated over the whole doping range only on ceramic samples.

A variety of bulk properties have already been studied in Tl2201. At optimal doping, the pure $d_{x^2-y^2}$ symmetry of the superconducting gap
was confirmed by phase-sensitive measurements \cite{Tsuei:1997}, and the so-called magnetic resonant mode at 47.5\,meV was detected below $T_c$
in neutron scattering experiments \cite{He:2002}, the first time for a single-layer cuprate. The high sample quality has been recently
demonstrated by measurements of angular magnetoresistance oscillations (AMRO), a bulk transport technique that requires the long mean free paths
afforded by high purity and crystallinity \cite{Hussey:2003}. The AMRO study of Tl2201 resulted in the first estimate of the normal state Fermi
surface of a cuprate superconductor from a bulk sensitive probe. The combination of charge \cite{Mackenzie:1996,Proust:2002} and heat
\cite{Proust:2002} transport measured at low temperature on very overdoped Tl2201 in the normal state, by applying up to 13T magnetic fields to
suppress superconductivity (see Fig.\,\ref{Tl-structure}), allowed the Lorenz ratio $L\!\equiv\!\kappa/\sigma T$ to be extracted. The value
obtained, $0.99 \pm 0.01\, L_0$, allowed one to conclude that the Wiedemann-Franz law is obeyed in overdoped cuprates:
\begin{figure}[t]
\begin{center}
\includegraphics[width=1\linewidth]{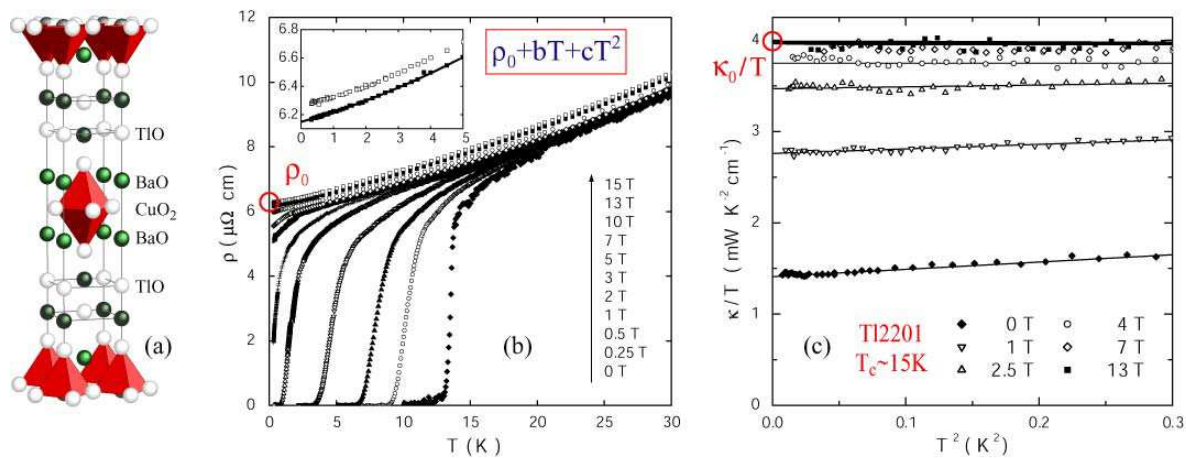}
\caption{\label{Tl-structure}(a) Unit cell of Tl2201. (b,c) Electrical resistivity and thermal conductivity of overdoped Tl2201
($T_c\!\simeq\!15\ K$), plotted vs.\ temperature for different values of an external magnetic field applied normal to the CuO$_2$ planes (after
Ref. \cite{Proust:2002}). From the $T\!=\!0$ K intercepts of $\rho(T)$ and $\kappa(T)/T$, the ratio $\rho_0\kappa_0/T\!=\!0.99\!\pm\!0.01 L_0$
was calculated ($L_0\!=\!2.44\times 10^{-8}\ W \Omega K^{_2}$ is Sommerfeld's value for the Lorenz ratio $L\!\equiv\!\kappa/\sigma T$), which
indicates that the Wiedemann-Franz law is obeyed in overdoped cuprates although the normal-state $\rho(T)$ is dominated by a non Fermi-liquid
like $T$-linear term (see inset).}
\end{center}
\end{figure}
the electronic carriers of heat are fermionic excitations of charge $-e$, suggesting that the low-energy excitations in overdoped cuprates might
indeed be described in terms of Fermi-liquid quasiparticles \cite{Proust:2002}. At variance with the hints of Fermi-liquid-like behavior,
however, the $ab$-plane resistivity is not purely described by a $T^2$ power law but is still dominated by a $T$-linear term (see inset of Fig
\ref{Tl-structure}). In this context, the detailed study of Tl2201 with modern single particle spectroscopies would be extremely desirable, as
it might provide direct insights into the nature of the quasiparticles in both the normal and superconducting states
 \cite{Andrea:2003,Campuzano,AndreaScripta}.

Unfortunately, despite its great potential, progress on the Tl2201 system has been slow because of difficulties in the growth of single
crystals. The thallium oxide is volatile, reactive, and highly toxic, a combination which places constraints on the crystal growth technique.
Standard growth procedures such as flux-growth in open crucibles or the use of a floating zone image furnace would allow toxic vapors to escape
into the lab and deplete the thallium in the reaction vessel by allowing it to sublime out in a colder part of the furnace. It is thus essential
to contain the growth components within a non-reactive enclosure so that thallium is not lost through either reaction or evaporation.

In Section \ref{preparation}, we report on our effort in the growth of high-purity single crystals of Tl2201, by a copper-rich self-flux method,
and in their careful annealing in a controlled oxygen atmosphere. We will show that this material can be cleaved to expose a clean and stable
[001] surface suitable for single particle spectroscopy experiments, as was demonstrated by the first successful study of the Fermi surface and
quasiparticle excitations of Tl2201 by ARPES \cite{mauro}. The detailed results of our ARPES study of several doping levels will be presented in
Section \ref{arpes}, and a discussion linked to other theoretical and experimental results will be given in Section \ref{discussion}. The
quantitative agreement between the ARPES and AMRO \cite{Hussey:2003} determined Fermi surfaces indicates that Tl2201 may be the ideal testing
ground for finally joining together the modern single particle spectroscopies and a host of well-established bulk probes of metals and
superconductors. In particular, Tl2201 might be the ideal cuprate for a broad-based study of the normal metal that may underlie the
superconductor in the overdoped regime, providing an understandable anchor point in the phase diagram of the high-temperature superconductors.

\section{Tl2201 Single-Crystal Preparation}
\label{preparation}

Difficult to prepare even as a ceramic, Tl2201 is even more challenging to grow as a crystal. Tl2201 melts incongruently and its crystals are
grown from the high temperature solution (melt) made of its components. Ideally, this melt would be decanted to reveal freestanding flux-free
crystals.  While this technique is commonly used for growth of complex oxide crystals, there are a few challenges particular to Tl2201 system.
The difficulties are due to the thallium oxide precursor. Around 800$^\circ$C, trivalent thallium oxide decomposes into a monovalent oxide and
oxygen gas \cite{Siegal}:
\begin{displaymath}
\mathrm{Tl}_2\mathrm{O}_3 \rightleftharpoons \mathrm{Tl}_2\mathrm{O} + \mathrm{O}_2\uparrow\,.
\end{displaymath}
While the oxygen gas evolved must be allowed to escape, this decomposition leads to a more serious problem: at growth temperatures, the
monovalent oxide's vapor pressure is of the order of 0.1\,atm, high enough to allow it to escape rapidly.  If the Tl$_2$O gas leaves the
crucible, however slowly, the composition of the melt will be at best time-varying and at worst quickly devoid of thallium. The volatility of
Tl$_2$O can also make the subsequent annealing of the crystals difficult, as the crystal surface may slowly decompose at temperatures at which
the oxygen atoms are mobile. There are several procedures that may be employed to mitigate these problems, including using pre-reduced thallium
precursors such as Tl$_2$Ba$_2$O$_5$ \cite{Jondo:1992}, encapsulation \cite{Hasegawa:2001}, or very high oxygen partial pressures.  A further
issue is that Tl$_2$O is quite reactive, most notably reacting with the quartz often used for encapsulation and furnace tubes.

To date, only a handful of groups have grown crystals of Tl2201.  Mackenzie's group, using a gold lid on their crucible as a barrier to
diffusion, found they could only successfully obtain crystals within a 2$^\circ$C by 2 minute window \cite{Tylerthesis,Liu:1992}. Hasegawa et
al.\ used reduced thallium precursors and an alumina high-pressure bomb to control thallium loss \cite{Hasegawa:1994,Hasegawa:2001}. Kolesnikov
et al. \cite{Kolesnikov:1995,Vyasilev:1992,Kolesnikov:1992} do not report any special precautions for controlling the loss of thallium.  In our
growth effort, as we will discuss in detail in Section \ref{growth}, we developed a flexible sealing scheme that permits the oxygen to escape as
it is evolved, while at the same time containing the Tl$_2$O.

A last crucial issue with the preparation of Tl2201 single crystals, alluded to above, is that the crystal surfaces can degrade during the
oxygen post-annealing required to set the charge carrier content, and in turn the temperature of the superconducting phase transition. So far
most work on annealing of Tl2201 has been carried out on ceramic samples \cite{Opagiste:1993}, but annealing of single crystals, where surface
damage is more obvious, has been less successful
--- only Mackenzie's group reported success \cite{Tylerthesis}. As a result, many crystals
studied to date have been unannealed, resulting in inhomogeneous oxygen content and broad transitions ($\Delta T_c\!=\!10\sim$20\,K), which can
obscure many features and add unnecessary disorder.  As part of our growth effort, we have also carefully explored the annealing of Tl2201
single crystals, succeeding in the preparation of samples ranging from near optimal doping to very overdoped, while avoiding surface damage.

\subsection{Single-Crystal Growth}
\label{growth}

For the crystal growth we employ a copper-rich self-flux technique using a barium and copper precursor, BaCuO$_2$ powder that was prepared from
99.999\% pure BaCO$_3$ and 99.995\% pure CuO by repeated calcining under flowing oxygen. The use of carbonate-free barium cuprate avoids both
carbon poisoning from BaCO$_3$ and the cation impurities that may be encountered if one employs BaO$_2$, which is only available at lower cation
purity. Carbon contamination is of particular concern in this system because there are several known superconducting thallium-based
oxycarbonates \cite{Raveau:1995}.

The barium cuprate powder was mixed with 99.99\% pure Tl$_2$O$_3$ to a final cation ratio of Tl:Ba:Cu = 1.05:1:1 \cite{Jorda}, then loaded into
a 10\,mL alumina crucible.
\begin{figure}[t]
\begin{center}
\includegraphics[width=0.93\textwidth]{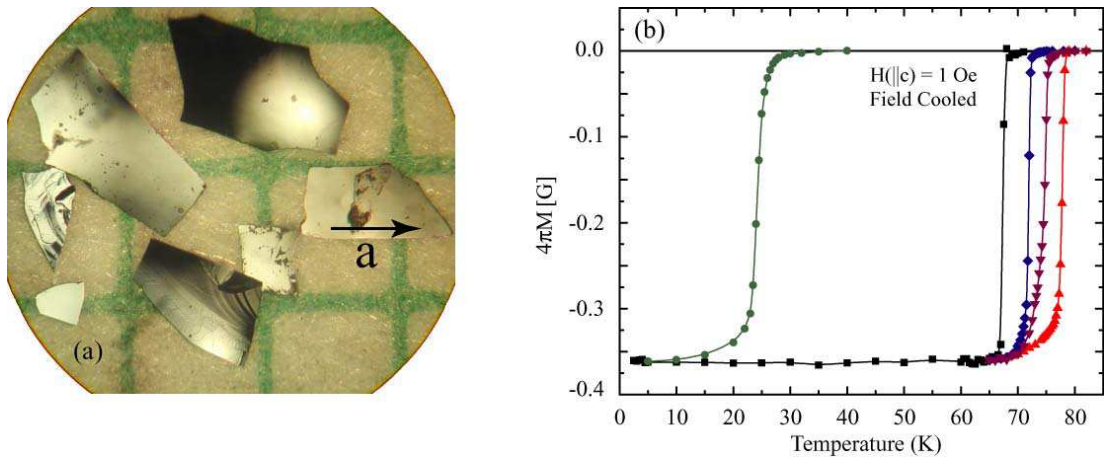}
\caption{\label{squid} (a) Photo of several resulting Tl2201 crystals on millimeter graph paper; the orthorhombic $a$-axis of one crystal is
marked. (b) Field-cooled magnetization curves for representative Tl2201 crystals annealed to different hole-doping levels; the sharpness of the
superconducting transitions is an indication of the crystals' homogeneity.}
\end{center}
\end{figure}
The crucible was sealed using a gold lid fixed in place with a 1\,kg weight. Employing this sealing scheme, O$_2$ gas may still escape as
Tl$_2$O$_3$ decomposes, but the seal becomes very effective once the contents of the crucible equilibrate at the growth temperature, thus
preventing significant loss of Tl$_2$O (as monitored by mass-loss measurements). The charge was heated rapidly (300\,$^\circ$C/h) to
935\,$^\circ$C, while flowing oxygen to flush out any escaping thallium oxide, and held for 12h to equilibrate. It was cooled at
$-0.5$$^\circ$C/h to 890\,$^\circ$C, then allowed to cool freely to room temperature. The oxygen gas flowing through the furnace was bubbled
through sulphuric acid and water to remove any thallium released during growth that did not condense in colder areas of the furnace. Most
resulting crystals were embedded in flux, but free-standing platelet single crystals of 1$\sim$2\,mm$^2$ were harvested from voids in the flux
ingot. As shown in Fig.\,\ref{squid}(a), these crystals exhibited mirror surfaces with fine curved growth steps. All characterization and
spectroscopy reported here was performed on flux-free, void-grown single crystals.

The as-grown samples typically have superconducting transitions of 5$\sim$10\,K (onset) and several Kelvins wide, a width comparable to other
groups' as-grown samples. As-grown $T_c$s are determined by the cation substitution level, crystal dimensions, cooling rate, and the atmosphere
around the crystals, the last two parameters being coupled through the equation of state for oxygen gas. Since as-grown superconducting
transitions are typically quite broad and not homogeneous throughout the crystal, it is crucial that the crystals be annealed before they are
studied. The crystals were annealed under controlled oxygen partial pressures at temperatures between 290\,$^\circ$C and 500\,$^\circ$C. This
produces samples whose oxygen contents place them on the overdoped side of the phase diagram, with sharp superconducting transitions ranging
from 5 to 85\,K. We have not attempted to reach optimal doping, which in this system is believed to correspond to a superconducting $T_c$ of at
least 93\,K \cite{Wagner}, because even higher annealing temperatures would be required, increasing the risk of thallium loss.

\subsection{Physical and Chemical Analysis}
\label{Characterization}

Since cation substitution can cause increased scattering, one objective of this work was to grow crystals with the lowest possible cation
substitution level. Of more immediate importance than the reduction of cation defects, however, is reproducibility.  Because cation substitution
also dopes the crystal, variation between crystals would lead to different $T_c$s as well as different scattering rates.  More problematic still
would be a variation within a crystal, which would arise from changes in the composition of the melt over the course of the crystal's growth.
Although cation substitution is not desired and does dope the crystals, it must be emphasized that this is not used as a control parameter in
doping these crystals.  If the crystals can be grown with a reproducible level of cation substitution, the cation disorder will be the same for
all dopings. This is in contrast to LSCO, where cation substitution is the primary control parameter, and must be varied by roughly a factor of
four to traverse the superconducting dome.

To determine the levels of cation substitution in our crystals, electron-probe micro-analyses (EPMA) of several crystals were performed on a
fully-automated CAMECA SX-50 instrument in wavelength-dispersion mode.\footnote{The EPMA was performed under the following operating conditions:
excitation voltage, 15\,kV; beam current, 20\,nA; peak count time, 80\,s; background count time, 40\,s;  and spot diameter, 10\,$\mu$m. Data
reduction was performed using the ``PAP'' $\varphi(\rho Z)$ method \cite{PAP}. For the elements studied, the following standards, X-ray lines
and monochromator crystals were used: elemental Tl, Tl{\slshape M}$\alpha$, PET; YBa$_2$Cu$_3$O$_{6.920}$, Ba{\slshape L}$\alpha$, PET; and
YBa$_2$Cu$_3$O$_{6.920}$, Cu{\slshape K}$\alpha$, LIF. Tight Pulse Height Analysis (PHA) control was used to eliminate to the degree possible
any interference from higher-order lines.}
\begin{figure}[t]
\begin{center}
\includegraphics[width=1\textwidth]{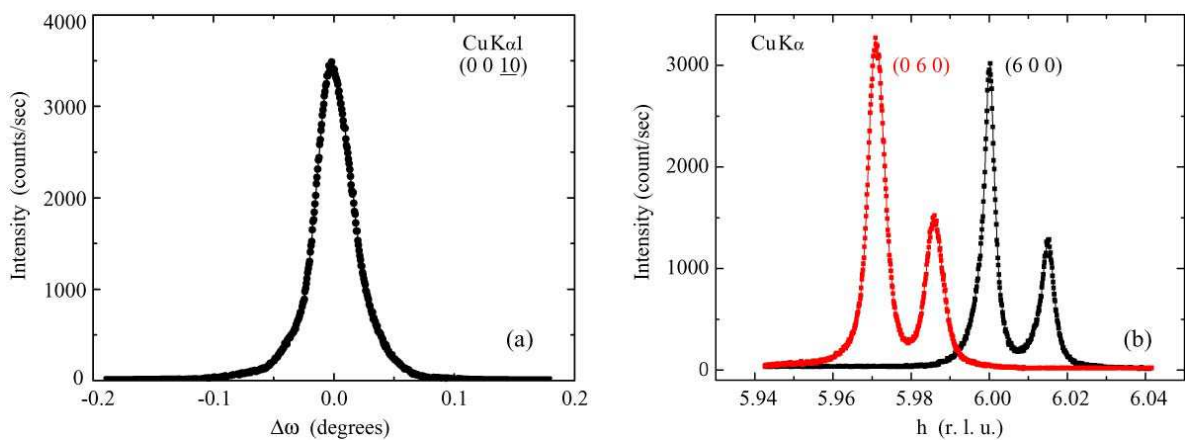}
\caption{\label{xrd}(a) The (0 0 10) rocking curve for a typical Tl2201 single crystal ({\slshape T}$_c$=67.7\,K); the FWHM is 0.034$^\circ$,
indicating excellent crystallinity. (b) The (6~0~0) and (0~6~0) peaks for a $T_c\!=\!24$\,K sample, showing the orthorhombicity of the crystal
structure. The doublets correspond to Cu$K\alpha_1$ and Cu$K\alpha_2$ radiation results.}
\end{center}
\end{figure}
The cation composition (normalized to the barium content) was determined to be Tl$_{1.884(6)}$Ba$_2$Cu$_{1.11(1)}$O$_{6+\delta}$ and was not
observed to be position dependent on individual crystals, nor did it vary between crystals.  This result corresponds to a level of copper
substitution at the thallium site comparable to that reported previously \cite{Shimakawa:1990,Tylerthesis,Liu:1992,Kolesnikov:1992} for
cation-substituted Tl2201. Al contamination was checked separately by EPMA and found to be below the 50\,ppm detection limit, corroborating our
observation that the crucibles were not corroded by the melt.\footnote{Operating conditions were the following:  excitation voltage, 15\,kV;
beam current, 40\,nA; peak count time, 500\,s; background count time, 250\,s;  and spot diameter, 10\,$\mu$m.  The matrix composition was fixed
at stoichiometric; Al{\slshape K}$\alpha$ was standardized on $\alpha$-Al$_2$O$_3$ using a TAP crystal monochromator.}
\begin{figure}[t]
\begin{center}
\includegraphics[width=0.58\textwidth]{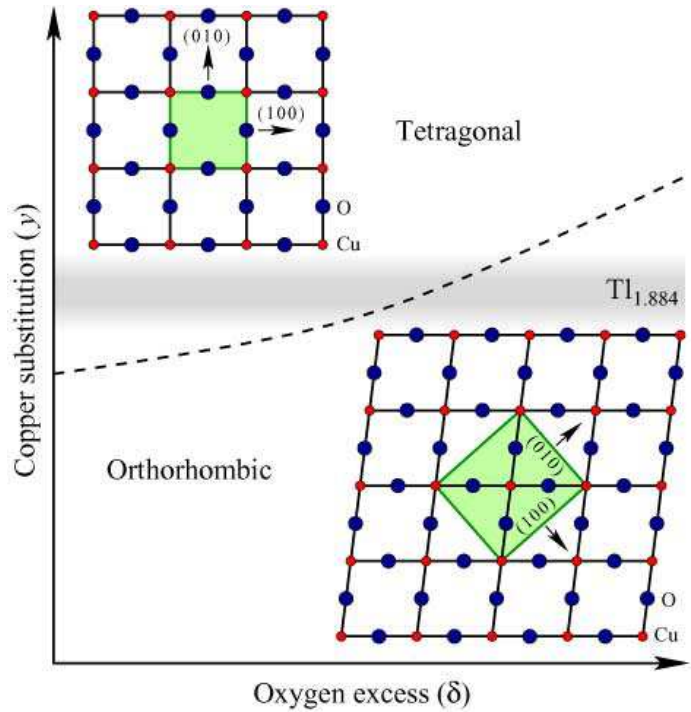}
\caption{\label{distortion}Schematic depiction of the dependence of the orthorhombic/tetragonal structural transition on $y$ and $\delta$ in
Tl$_{2-y}$Ba$_2$Cu$_{1+y}$O$_{6+\delta}$. This scenario is suggested by the data from polycrystalline samples of Wagner et al. \cite{Wagner} and
our own single crystal diffraction results (see Fig.\,\ref{lattice}). The CuO$_2$ plane and unit cell (in green) are shown with and without the
orthorhombic distortion, which is here exaggerated for clarity. The gray bar indicates where our crystals fall on this plot:  $y$=0.116.}
\end{center}
\end{figure}
Despite the cation non-stoichiometry, the high quality of these crystals is evidenced by their narrow superconducting transitions, measured by
magnetization in a Quantum Design SQUID magnetometer (MPMS). Fig.\,\ref{squid}(b) shows the field-cooled magnetization measured at 1\,Oe
($H\parallel c$), for crystals with transition widths (10\%-90\%) ranging from 4 to 0.7\,K. The observed 40$\sim$60\% Meissner fraction (i.e.,
fraction of flux excluded when cooling through {\slshape T}$_c$) compares well with field-cooled results on other superconducting cuprates. It
should also be mentioned that Tl2201's extremely low first critical field ($H_{c1}$) makes the transitions appear broader in higher fields.

\subsection{Structural Analysis: Crystallinity and Symmetry}

To investigate the samples' crystallinity, X-ray diffraction spectra and rocking curves were taken on a BEDE model 200 double-crystal
diffractometer with a Si(111) monochromator;  the entire sample was illuminated by the X-ray beam.  Fig.\,\ref{xrd}(a) depicts the (0 0 10)
X-ray rocking curve of a 0.5$\times$1.2\,mm$^2$ crystal annealed to {\slshape T}$_c$=67.7\,K. While the FWHM of 0.034$^\circ$ is broader than
the 0.006$^\circ$ obtained for the best YBa$_2$Cu$_3$O$_{7-\delta}$ crystals grown in BaZrO$_3$ crucibles, it is similar to the
0.02-0.03$^\circ$ achieved for YBa$_2$Cu$_3$O$_{7-\delta}$ grown in YSZ crucibles \cite{Ruixing:1998}, indicating a good degree of crystalline
perfection.
\begin{figure}[t]
\begin{center}
\includegraphics[width=1\textwidth]{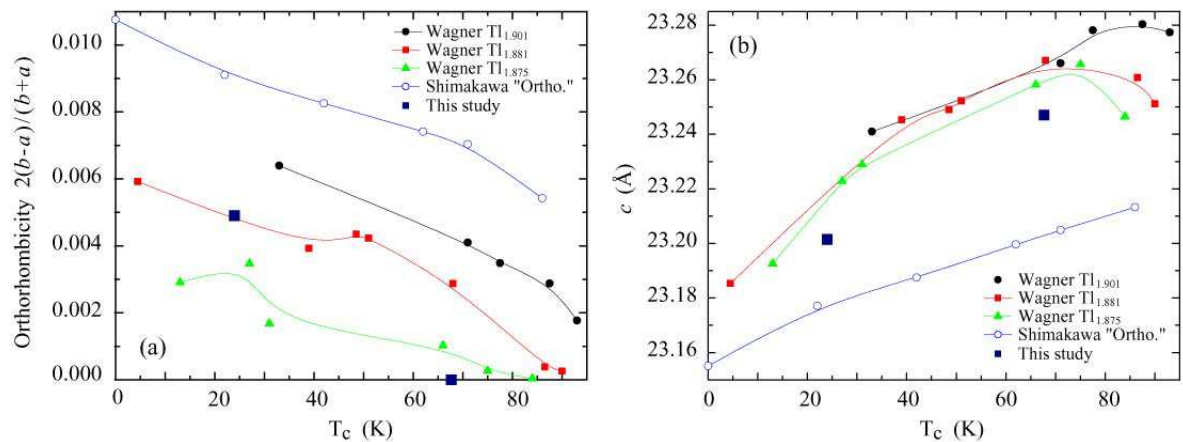}
\caption[Lattice parameters]{\label{lattice}Comparison of the orthorhombicity between our crystals and Wagner's \cite{Wagner} and Shimakawa's
\cite{shimakawa:1993} ceramics: orthorhombicity and $c$-axis lattice constant vs.\ $T_c$.}
\end{center}
\end{figure}
A crystal annealed to achieve a relatively high oxygen content and strong overdoping ({\slshape T}$_c$=24\,K) was found to be orthorhombic with
lattice parameters $a$=5.4580(3)\,\AA, $b$=5.4848(5)\,\AA\, and $c$=23.2014(5)\,\AA, consistent with the lattice constants reported for ceramics
with similar {\slshape T}$_c$ and Cu substitution \cite{Wagner}. This orthorhombicity is determined by the x-ray diffraction data presented in
Fig.\,\ref{xrd}(b): the two data sets were recorded independently for $a$ and $b$ axes and would be identical to one another if the crystals
were tetragonal. Closer to optimal doping, and at a lower interstitial oxygen content, the {\slshape T}$_c$=67.7\,K sample was tetragonal to
within our resolution

Tl2201's crystal symmetry depends on the level of cation substitution and the oxygen content. Stoichiometric and near-stoichiometric ceramics
have been shown to be orthorhombic for all oxygen contents, while heavily-substituted ones are strictly tetragonal \cite{shimakawa:1993}. For
intermediate levels of substitution, the orthorhombicity depends on the oxygen content: samples with higher oxygen contents are more
orthorhombic \cite{Wagner}.  The orthorhombic distortion is thought to arise from a lattice mismatch between the CuO$_2$ plane layer and the
naturally larger Tl$_2$O$_2$ double layer. Interstitial oxygens increase the distortion, while cation disorder seems to suppress it. The
distortion and its dependence on the oxygen content $\delta$ and the copper substitution $y$ are shown schematically in Fig.\,\ref{distortion}.
Powder X-ray diffraction shows that the orthorhombic distortion is an elongation along one plaquette diagonal, with the Cu-O bonds remaining the
same in both directions. The distortion is quite minor: instead of being at right angles, the Cu-O bonds meet at $\approx$89.7$^\circ$ in
orthorhombic samples. The plaquette diagonals, identified with nodes in the superconducting gap, remain orthogonal, so no mixing of order
parameter symmetries is required.  For these reasons and for ease of comparison with other systems, all ARPES analysis reported here is based on
the original tetragonal unit cell, regardless of the actual symmetry of the crystals.

Our crystals' orthorhombicity, defined as $2(b-a)/(b+a)$, is consistent with that found by Wagner et al. \cite{Wagner} for similar levels of
cation substitution, as shown in Fig.\,\ref{lattice}(a). One should keep in mind, however, that this quantity is highly prone to uncertainties
because of the smallness of the orthorhombic distortion, so that quantitative conclusions based on this graph should be treated with caution.
Fig.\,\ref{lattice}(b) compares two of our annealed crystals' $c$-axis lattice parameters to the orthorhombic ceramic samples of Wagner et al.
\cite{Wagner} and Shimakawa et al. \cite{shimakawa:1993}. Our $c$-axis lattice parameters are again similar to those of Wagner et al.
\cite{Wagner}, but we observe higher $T_c$ values for the same lattice parameters (or a shorter $c$ axis for the same $T_c$). Possible
explanations for this include improvements in $T_c$ from higher homogeneity, cleaner samples or larger grain size (crystals vs.\ ceramics), more
homogeneous annealing, or possibly a calibration disagreement between the diffractometers. Wagner's samples have transitions a few Kelvins wide,
with long tails towards zero Kelvin, and $T_c$ was defined using the 50\% point, which may not be an accurate average of the bulk in such a
situation.

\section{The Low-Energy Electronic Structure of Tl2201}
\label{arpes}

Angle-resolved photoemission spectroscopy (ARPES) is an important tool in the study of the electron dynamics in correlated-electron systems.
Within the sudden approximation, it probes the energy and momentum dependence of the electronic excitation spectrum of an $N-1$ particle system,
the so-called electron-removal portion of the single-particle spectral function $A({\bf k},\omega)$ \cite{AndreaScripta}. In the non-interacting
picture, this spectral function consists of delta-function peaks located at the precise energy and momentum given by the band structure, i.e.
$A({\bf k},\omega)\!=\!\delta(\omega\!-\!\epsilon_{\bf k})$. When interactions are considered, the single-particle spectral function is modified
by the inclusion of the electron proper self energy $\Sigma({\bf k},\omega)\!=\!\Sigma^{\prime}({\bf k},\omega)\!+\!i\Sigma^{\prime\prime}({\bf
k},\omega)$, which captures all the of the many-body correlation effects. One can then write $A({\bf
k},\omega)=-\frac{1}{\pi}\frac{\Sigma^{\prime\prime}({\bf k},\omega)}{[\omega-\epsilon_{\bf k}-\Sigma^{\prime}({\bf
k},\omega)]^2+[\Sigma^{\prime\prime}({\bf k},\omega)]^2}$. With respect to the non-interacting case, the peaks in the spectral function shift in
energy and gain a finite width, in a way dependent on the energy and momentum of the excitations. At those $\omega$ and $k$ for which the
spectral function is still characterized by a single pole, energy and lifetime renormalization are directly described by $\Sigma^{\prime}({\bf
k},\omega)$ and $\Sigma^{\prime\prime}({\bf k},\omega)$, respectively. The ARPES lineshape thus gives direct access to the lifetime of the
excitation and can provide insights into the nature of the underlying interactions, for example whether or not electron-electron interactions
are Fermi liquid-like, as discussed in relation to the Tl2201 transport data of Fig.\,\ref{Tl-structure}. As it will be discussed in the
following sections, the first successful ARPES experiments on Tl2201 have arrived at an agreement with bulk probes on key features such as the
normal-state Fermi surface \cite{Hussey:2003} and the superconducting gap \cite{Hawthorn}. This success suggests that detailed ARPES studies of
Tl2201 have the potential to reveal the nature and strength of many-body correlations, upon approaching high-$T_c$ superconductivity from the
more conventional overdoped regime.

\subsection{Band Structure Calculations}

Since it is generally believed that the normal metal on the very overdoped side of the HTSC phase diagram can be described as a rather
conventional Fermi liquid system, in contrast to the strongly correlated Mott insulator found at half filling ($p\!=\!0$ in
Fig.\,\ref{doping-diagram}), exploring the electronic structure and Fermi surface of heavily overdoped Tl2201 can start with the results of
non-interacting band structure calculations performed within the local density approximation (LDA).
\begin{figure}[t]
\begin{center}
\includegraphics[width=4in]{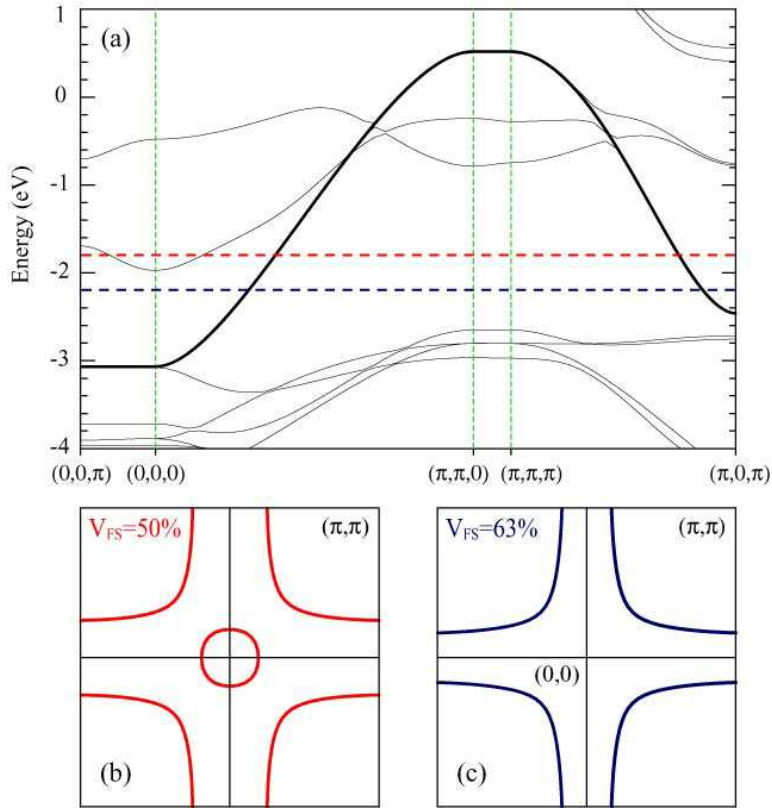}
\caption{\label{LDA} (a) Electronic dispersion obtained from band structure calculations within the local-density approximation (LDA). (b,c)
Fermi surface calculated for two different doping levels enclosing a volume, counting holes, of 50\% (orange) and 63\% (blue) of the
two-dimensional projected Brillouin zone; the corresponding position of the chemical potential with respect to the dispersive electronic bands
is indicated in panel (a). Here, as in the rest of the paper, $k$-space labels are expressed modulo the lattice constants.}
\end{center}
\end{figure}
As shown in Fig.\,\ref{Tl-structure} and especially Fig.\,\ref{distortion}, the most important structural element is the CuO$_2$ planes, which
are well separated from each other by the TlO and BaO layers. As in all other HTSC cuprates, the CuO$_2$-derived bands are expected to be the
lowest energy electronic states and therefore directly determine the macroscopic electronic properties.

The results of band structure calculations along the high symmetry directions in the body-centered tetragonal Brillouin zone of Tl2201 are
presented in Fig.\,\ref{LDA}(a). Coordinates for the atoms are taken from Ref.\,\cite{Singh:1992}.  The essential low-energy feature of the band
structure is indeed the highly dispersive Cu$(3d_{x^2-y^2})$-O(1)$(2p_{x,y})$ band; note however that this band is highly two-dimensional, with
little dispersion along the $k_z$ direction $(0,0,k_z\!=\!0)\! \rightarrow \!(0,0,k_z\!=\!\pi)$. Its top and bottom are located at ($\pi,\pi$)
and $(0,0)$, respectively, within the two-dimensional projected Brillouin zone, while at $(\pi,0)$ we find the well-known van Hove singularity.
The other two bands that sit at a comparable energy, i.e 0 to $-\!2$\,eV in Fig.\,\ref{LDA}(a), are the anti-bonding Tl(6s)-O(2),O(3)(p$_z$)
bands, which do show significant dispersion in the $k_z$ direction. Based on the formal valences of stoichiometric and undoped
Tl$_2$Ba$_2$CuO$_{6+\delta}$ (2:2:1:6 with Tl$^{3+}$, Ba$^{2+}$, Cu$^{2+}$, O$^{2-}$, $\delta\!=\!0$), the Fermi energy would be located at the
red line in Fig.\,\ref{LDA}(a), and both the CuO and TlO bands would cross E$_f$. This would result in a Fermi surface with hole pockets
associated with the CuO band, and a small, spheroidal TlO electron pocket at the $\Gamma\!=\!(0,0)$ point, as in Fig.\,\ref{LDA}(b). This small
electron pocket was originally proposed as a possible explanation for why undoped Tl$_2$Ba$_2$CuO$_{6}$ does not show the Mott insulating
behavior expected for materials with a half-filled $3d_{x^2-y^2}$ CuO band, such as undoped La$_2$CuO$_4$ \cite{Singh:1992}. However, the
non-stoichiometry of our samples (Tl$_{1.884(6)}$Ba$_2$Cu$_{1.11(1)}$O$_{6+\delta}$, see Section \ref{Characterization}) provides additional
hole doping ($\sim\!0.14$ holes/formula unit) that would push the CuO band further away from half-filling, driving the TlO band above E$_f$.
This shift would generate a Fermi surface consisting solely of a CuO hole pocket around ($\pi,\pi$) similar to the results shown in blue in
Fig.\,\ref{LDA}(a,c), which were calculated to match the 63\% volume observed by ARPES on our $T_c\!=\!30$ K overdoped Tl2201 crystal
\cite{mauro}. Finally, it should be mentioned that if the effect of hole doping through interstitial oxygen were explicitly included in the
calculations, the TlO bands would be pushed to much higher energies, beyond the rigid band picture \cite{SahrakorpiTl2201}. The analogous
effects of Pb substitution or excess oxygen in the Bi-O layers of Bi-cuprates has recently been used to account for the lack of Bi-O electron
pockets around ($\pi$,0) in those materials \cite{lin-xxx}.

\subsection{Electronic Dispersion and Fermi Surface by ARPES}

\begin{figure}[t]
\begin{center}
\includegraphics[width=5.5in]{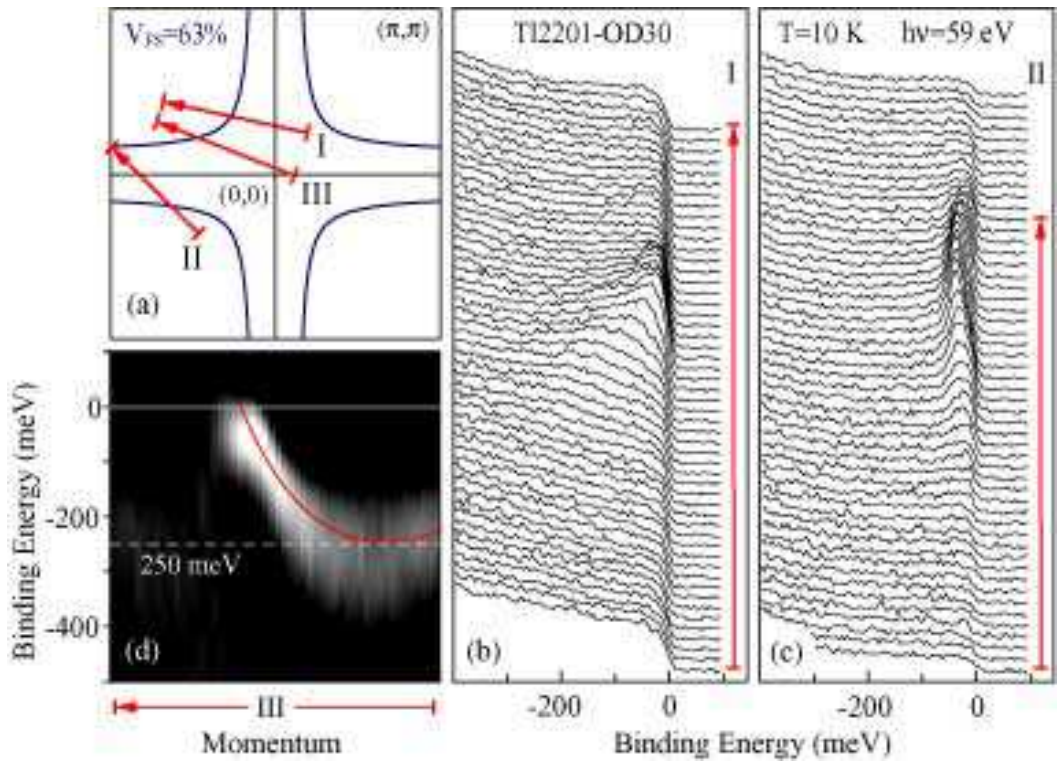}
\caption{\label{Tl-QP} (a) Fermi surface calculated for a 63\% Brillouin zone volume, counting holes, as in Fig.\,\ref{LDA}(c). (b,c) ARPES
spectra taken at $T\!=\!10$\,K on Tl2201-OD30 along the directions marked by arrows I and II in (a). (d) Second derivative vs.\ energy of the
 spectra from along arrow III in (a); the red line is our tight-binding fit (see text).}
\end{center}
\end{figure}

Fig.\,\ref{Tl-QP}(b) shows ARPES data\footnote{The ARPES experiments were carried out at the Swiss Light Source (SLS) on the Surface and
Interface Spectroscopy Beamline \cite{19a,19b,19c} and at the Stanford Synchrotron Radiation Laboratory (SSRL) on Beamline 5-4, in both cases
using a Scienta SES-2002 photoelectron spectrometer.  At SLS, all measurements were performed with circularly polarized 59\,eV photons, with
energy and angular resolutions of approximately 24\,meV and 0.2$^\circ$; the data were acquired on a strongly overdoped sample with {\slshape
T}$_c$=30\,K (Tl2201-OD30), and a lightly overdoped sample with {\slshape T}$_c$=63\,K (Tl2201-OD63). The samples were cleaved {\it in situ} at
$6\!\times\!10^{-11}$\,torr and kept at 10\,K throughout the measurements.  At SSRL, all data were acquired at 28\,eV with linearly polarized
light and with energy and angular resolutions of 15\,meV and 0.35$^\circ$.  The SSRL data were from lightly overdoped samples with {\slshape
T}$_c$=74\,K (Tl2201-OD74), which were cleaved at $3\!\times\!10^{-11}$\,torr and temperature cycled between 10 and 85\,K.}
from a $T_c\!=\!30$\,K very overdoped Tl2201 sample (Tl2201-OD30) taken close to the (0,0)-($\pi$,$\pi$) direction in momentum space, the
so-called ``nodal'' direction where the $d$-wave superconducting gap is zero.
Each line in Fig.\,\ref{Tl-QP}(b) is an energy distribution curve (EDC) from a given position in momentum space along the line marked as
$\textrm{I}$ in the Fermi surface plot of Fig.\,\ref{Tl-QP}(a). These EDCs show a strongly dispersing quasiparticle peak, related to the
$3d_{x^2-y^2}$ CuO band, which emerges from the background at high binding energies and progressively sharpens as it disperses all the way to
the Fermi energy $E_F$. Note that the term {\it quasiparticle} (QP) is used in a loose sense to identify a reasonably sharp dispersive peak,
without specifying in detail how the peak should be interpreted in terms of specific models for correlation effects (more discussion will be
given in Sections \ref{lineshape} and \ref{discussion}). From the results of band structure calculations (Fig.\,\ref{LDA}), we expect the bottom
of this band to be located at the $\Gamma$ point, or (0,0), at a binding energy of approximately 1\,eV with respect to the chemical potential.
Although it is hard to track the band all the way down to high binding energies in the raw data, it is possible to highlight the band dispersion
by taking a second derivative of the ARPES intensity with respect to energy. This is plotted in Fig.\,\ref{Tl-QP}d and indicates a filled
bandwidth of only $\sim\!250$\,meV, suggesting a renormalization factor of about 4 between single particle band structure calculations and
measured ARPES spectra.

Near ($\pi$,0), the so-called ``antinodal'' direction where the $d$-wave superconducting gap
exhibits its largest value, we also detected well-defined dispersive QP peaks, which define a
shallow parabolic band. The bottom of this band corresponds to the extended van Hove
singularity observed near ($\pi$,0) in all superconducting cuprates
\cite{Andrea:2003,Campuzano}.
\begin{figure}[t]
\begin{center}
\includegraphics[width=1\linewidth]{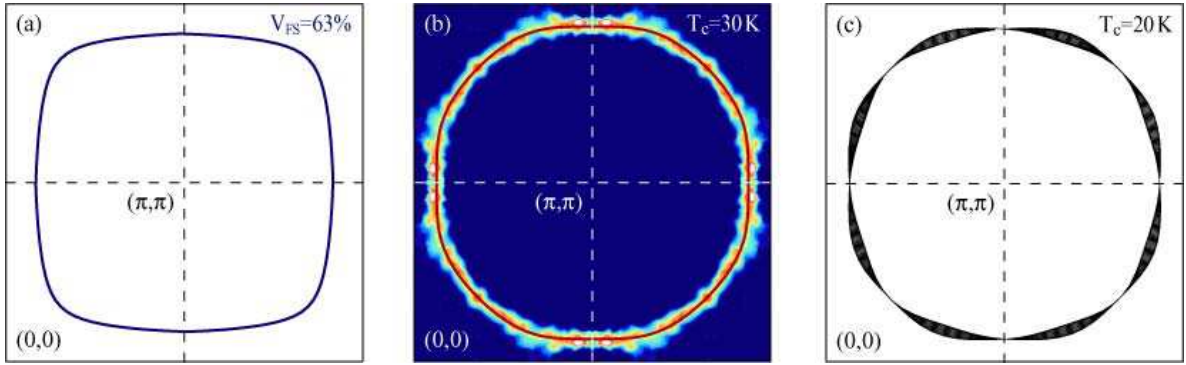}
\caption{\label{FS} Fermi surfaces, in the projected two-dimensional Brillouin zone, obtained from: (a) band structure calculations (imposing a
volume of 63\% counting holes, as observed by ARPES on Tl2201-OD30); our (b) ARPES experiments on the $T_c\!=\!30$\,K sample (Tl2201-OD30); and
(c) the AMRO study on a $T_c\!=\!20$\,K Tl2201 crystal by Hussey et al. \cite{Hussey:2003}. The red line in (b) is the result of our
tight-binding fit of ARPES Fermi surface and dispersion; the width of the Fermi surface contour in (c) reflects the magnitude of the $c$-axis
dispersion, here emphasized by a factor of 4 for clarity \cite{Hussey:2003}.}
\end{center}
\end{figure}
However, at this high doping level, the van Hove singularity is located approximately 40\,meV below the chemical potential. In comparison to the
band structure calculations, this also shows a renormalization factor of about 5.

The momentum at which a dispersing QP peak is observed to reach $E_F$ and disappear provides ARPES's means of determining a Fermi wavevector
$k_F$. This wavevector identifies one point along the normal state Fermi surface, which is the contour in momentum space that separates filled
from empty electronic states and whose existence and details are of fundamental importance for the understanding of the macroscopic physical
properties of a material.  By integrating the ARPES spectra over a $\pm$5\,meV energy window about $E_F$ for a large number of cuts in momentum
space and then plotting the results versus momentum in the two-dimensional projected tetragonal Brillouin zone, one obtains an estimate of the
normal-state Fermi surface.  This has been done for Tl2201-OD30, and is shown in Fig.\,\ref{FS}(b). The location of the Fermi surface crossings
has been determined over more than one quadrant across two different zones, and then downfolded to the first Brillouin zone and four-fold
symmetrized to clearly show the detailed shape of the Fermi surface in the reduced zone scheme, with improved signal-to-noise. As expected,
there is no indication of a TlO electron pocket around (0,0). The Fermi surface determined from this procedure takes the form of a large hole
pocket centered at ($\pi$,$\pi$) with an area that occupies 63$\pm$2\% of the Brillouin zone, corresponding to a carrier concentration of
1.26$\pm$0.04 hole/Cu atom, or $p$=0.26$\pm$0.04 greater hole density than the half-filled Mott insulator with 1 hole/Cu ($p\!=\!0$ in
Fig.\,\ref{doping-diagram}). This result is in superb agreement with the recent study of the Fermi surface by angular dependence of
magnetoresistance oscillations (AMRO) experiments \cite{Hussey:2003}, which found a hole-pocket volume of 62\% of the Brillouin zone in a
slightly more overdoped {\slshape T}$_c\!=\!20$\,K sample (Fig.\,\ref{FS}(a)). The ARPES determination is also in good agreement with the
estimates from low temperature measurements of the Hall coefficient, which gave a hole doping of $p\!=\!1.30$ holes/Cu for a
$T_c\!\lesssim\!15$\,K sample \cite{Mackenzie:1996}. It is worth noting that the commonly used empirical formula,
$T_c/T_c^{max}\!=\!1\!-\!82.6(p\!-\!0.16)^2$, which purports to connect the value of $T_c$ to doping in a universal fashion \cite{Presland},
gives a much stronger dependence of $T_c$ on doping than is shown by ARPES, AMRO, and Hall coefficient data. For our Tl2201-OD30 sample, from
the above formula and the value $p\!=\!0.26$ given by the Fermi surface volume, one would obtain $T_c\!\simeq\!16$\,K; and for the
$T_c\!=\!15$\,K sample studied by Mackenzie et al. \cite{Mackenzie:1996}, with the formula and the value $p\!=\!0.30$ determined from the Hall
coefficient, one would actually get $T_c\!=\!0$\,K. According to the empirical formula and contrary to what observed, Tl2201 should become
non-superconducting already for $p\!\simeq\!0.27$.

As a quantitative measure of the shape of the Fermi surface and of the many-body renormalized electronic dispersion, the Tl2201-OD30 ARPES data
can be modelled by the tight-binding formula $\epsilon_{\bf k}\!=\!\mu\!+\!\frac{t_1}{2}(\cos k_x\!+\!\cos k_y)\!+\!t_2\cos k_x\cos
k_y\!+\!\frac{t_3}{2}(\cos 2k_x\!+\!\cos 2k_y)\!+\!\frac{t_4}{2}(\cos 2k_x\cos k_y\!+\!\cos k_x\cos 2k_y)\!+\!t_5\cos 2k_x\cos 2k_y$
\cite{Norman:2001}, setting $a\!=\!1$ for the lattice constant. With parameters $\mu\!=\!0.2438$, $t_1\!=\!-0.725$, $t_2\!=\!0.302$,
$t_3\!=\!0.0159$, $t_4\!=\!-0.0805$, $t_5\!=\!0.0034$, all expressed in eV, this dispersion reproduces both the Fermi surface shape, solid red
line in Fig.\,\ref{FS}(b), and the QP energy at (0,0) and ($\pi$,0), as seen in Fig.\,\ref{Tl-QP}(c,d).  It is worth noting that experimentally
the band bottom at ($\pi$,0) is extremely flat, a behavior that could not be reproduced by including only $t_1$ and $t_2$ hopping parameters in
the model. Alternatively, a simple analytical formula for the three-dimensional electronic dispersion of Tl2201 has recently been derived,
within the framework of the linear combination of atomic orbitals (LCAO) approximation, and was used to fit both ARPES and AMRO results
\cite{mishonov}. A basis set spanning the Cu $4s$, Cu 3$d_{x^2-y^2}$, O $2p_x$ and O $2p_y$ states was used in order to take into account the
effective Cu-Cu hopping amplitude between Cu $4s$ orbitals in neighboring CuO$_2$ layers. This approach emphasizes the significance of the Cu
$4s$ hopping amplitude in order to obtain the correct three-dimensional Fermi surface shape, an issue that was originally raised by Andersen
{\em et al.} \cite{andersen}. With regards to the detailed shape of the Fermi surface, LDA band structure calculations predict a more square
contour than observed by ARPES and AMRO (Fig.\,\ref{FS}(a)). The inclusion of correlation effects, as well as Cu-Tl substitution or interstitial
O-doping beyond a rigid-band picture, might lead to a better agreement; preliminary attempts in this direction have not yet led to qualitative
improvements \cite{SahrakorpiTl2201}.

\subsection{ARPES Study of the Superconducting Gap}

In order for Tl2201 to be a model system to study the overdoped HTSCs by surface and bulk sensitive probes, it is important to show that ARPES
measures quantities that are characteristic of the bulk, for both normal and superconducting states. Therefore, in addition to the quantitative
agreement on the normal state Fermi surface, the observance of a superconducting gap in agreement with bulk measurements is also a necessary
requirement. In the following, we will discuss three different methods for the observation of a gap by ARPES.
\begin{figure}[t]
\begin{center}
\includegraphics[width=1\textwidth]{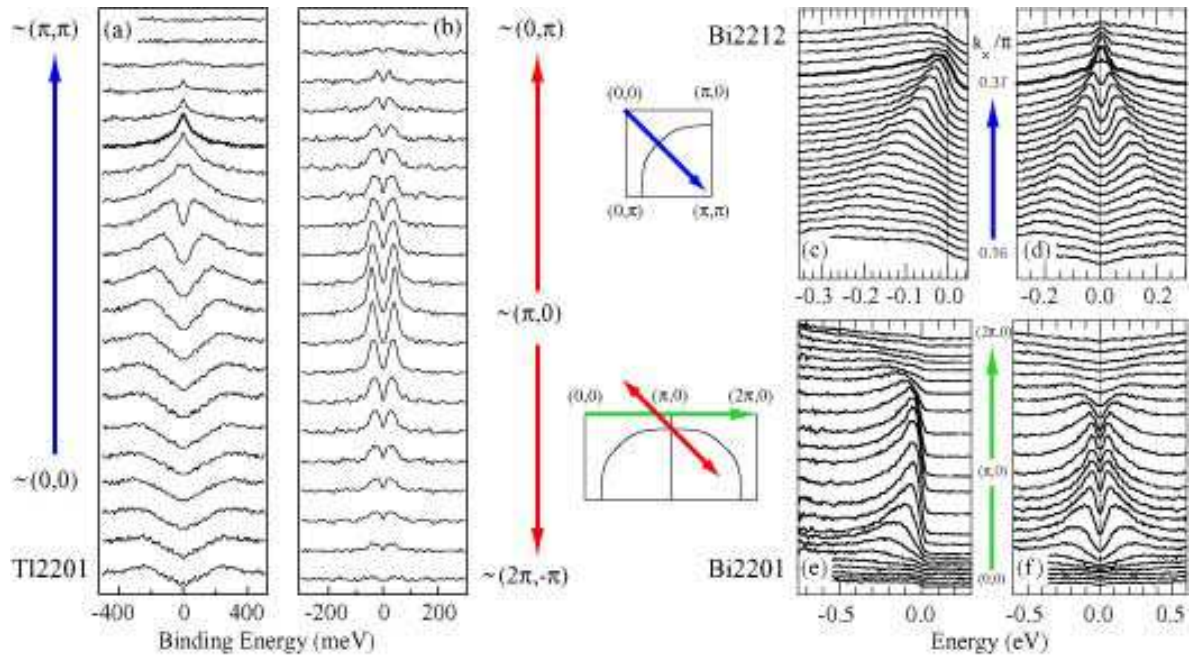}
\caption{\label{Tl-gap-symmetrized}(a,b) Symmetrization of Tl2201-OD30 ARPES spectra from along cuts I and II in Fig.\,\ref{Tl-QP}(b,c). ARPES
spectra and their symmetrization along (c,d) the nodal direction for overdoped Bi$_2$Sr$_2$CaCu$_2$O$_{8+\delta}$ ($T_c\!=\!88$\,K), and (e,f)
perpendicular to the antinodal direction for overdoped Bi$_2$Sr$_2$CuO$_{6+\delta}$ ($T_c\!=\!23$\,K); after Ref.\,\cite{Mesot:2001}. For all
samples, the approximate locations of the $k$-space cuts are indicated in the Brillouin zone sketches. Bold lines in (a,c,d) mark the spectra
that cross the Fermi energy, identifying a Fermi wavevector $k_F$; no crossing is observed for the spectra in (b,e,f).}
\end{center}
\end{figure}
The first two show a gap consistent with a $d_{x^2-y^2}$ form. The third method allows one to follow the temperature dependence of the gap,
highlighting the minimal surface degradation that occurs as the temperature is cycled from 10 to 85\,K.

The detection of a $d_{x^2-y^2}$ gap using ARPES can be most easily visualized by the comparison of nodal and antinodal symmetrized spectra
\cite{Norman:1998}.  The spectra are symmetrized in energy about E$_f$, by taking $I(\omega)\!+\!I(-\omega)$, which minimizes the effects of the
Fermi function.\footnote{The symmetrization procedure assumes that the ARPES spectra are described by $I(\omega)\!=\!f(\omega)A({\bf
k},\omega)$, and that there is particle-hole symmetry for a small range of $\omega$ about $E_F$ such that $A(-\epsilon_{\bf
k},-\omega)\!=\!A(\epsilon_{\bf k},\omega)$.  With the identity $f(-\omega)\!=\!1\!-\!f(\omega)$, it then follows that
$I(\omega)\!+\!I(-\omega)\!=\!A({\bf k},\omega)$. It is worth pointing out that this procedure is not strictly valid in the case that energy and
momentum resolutions are included in the description of $I(\omega)$, via convolution with the resolution function $R(\Delta k,\Delta\omega)$.}
While this procedure does not return a quantitative value for the size of the superconducting gap, it provides a qualitative criterion for
determining whether or not there is a Fermi crossing, and hence whether or not a superconducting gap has opened along the normal-state Fermi
surface.
\begin{figure}[t]
\begin{center}
\includegraphics[width=1\textwidth]{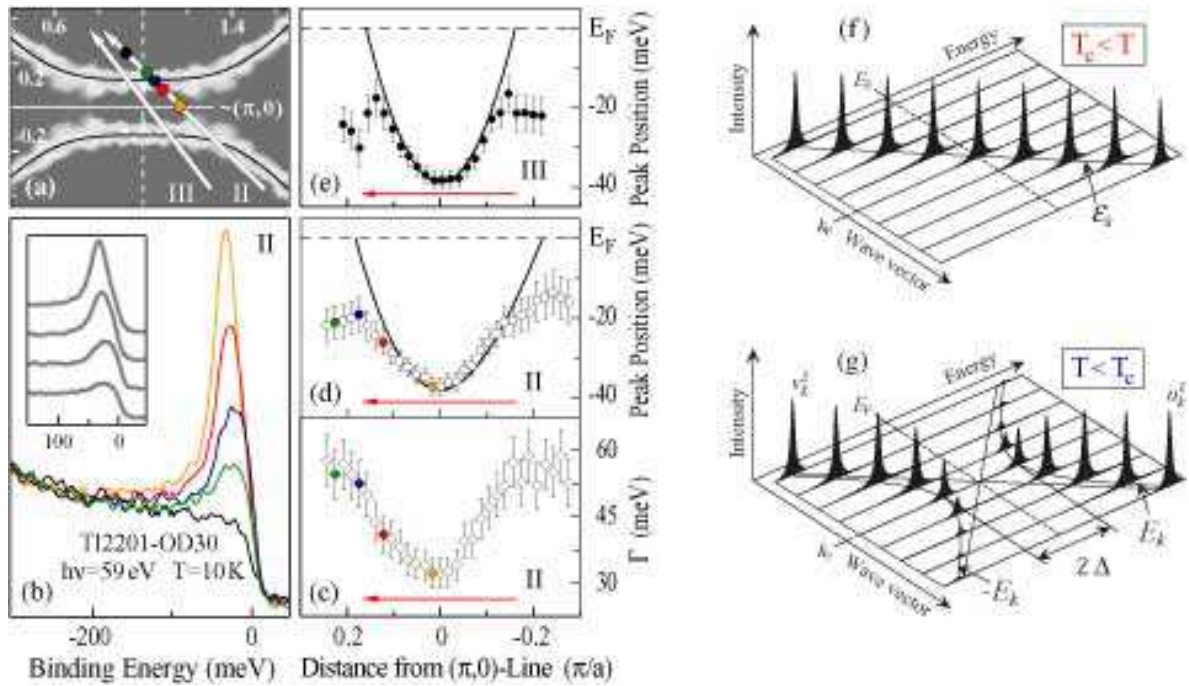}
\caption{ \label{Tl-gap-fitted} (a) Enlarged view of the Fermi surface of Tl2201-OD30 near ($\pi$,0). (b) Selected spectra from along cut II in
(a); their $k$-space positions are indicated by circles of corresponding color. (c,d) QP linewidth $\Gamma$ and peak position from a Lorentzian
fit of the energy distribution curves along cut II in (a). (e) Similarly, QP peak position along cut III in (a). Black lines in (a,d,e) are our
tight-binding results (see text). (f) Normal and (g) superconducting state single-particle spectral function, highlighting particle-hole mixing
and backward dispersion of Bogoliubov QPs below $T_c$ (after Ref.\,\cite{Matsui:2003}).}
\end{center}
\end{figure}
In the symmetrized ARPES data, the presence of a peak in the spectra at $E_F$ indicates the presence of a Fermi surface crossing. This procedure
has been used extensively for the Bi-based cuprates, both single and bilayer compounds, in detailed investigations of the normal-state Fermi
surface, as shown for instance in Fig.\,\ref{Tl-gap-symmetrized}(c-f) \cite{Mesot:2001}, and of the superconducting as well as normal state
pseudogap \cite{Norman:1998,Kanigel:2006}. In the case of our Tl2201-OD30 sample, this crossing is clearly seen in the nodal direction (bold
line in Fig.\,\ref{Tl-gap-symmetrized}(a)) but not in the antinodal direction (Fig.\,\ref{Tl-gap-symmetrized}(b)), which is consistent with a
$d$-wave functional form for the gap.

A more quantitative analysis of the gap can be performed by fitting ARPES spectra along cuts that cross the underlying normal-state Fermi
surface, as shown in Fig.\,\ref{Tl-gap-fitted}. The model used to reproduce the spectra is a Lorentzian peak plus a step-like background. The
latter is determined from the ARPES spectra with $k\gg k_F$ and is used to help phenomenologically isolate the coherent part of the spectral
function. This function is then multiplied by the Fermi function and convolved with the instrumental energy resolution function, to obtain the
functional form to be fit to the data \cite{Andrea:2003}. This procedure for determining the gap has been used previously for
Bi$_2$Sr$_2$CaCu$_2$O$_{8+\delta}$ \cite{Ding:1996}. The inset of Fig.\,\ref{Tl-gap-fitted}(b) shows good agreement between the raw data and the
fit. Since heavily overdoped cuprates have weaker, possibly Fermi liquid-like electron correlations, good agreement between the measured QP peak
and the Lorentzian is in principle expected.

Fig. \ref{Tl-gap-fitted}(d,e) show the peak positions from the fits compared with our tight-binding description of the normal-state electronic
structure near the antinodal region. At higher binding energies, there is good agreement between the fit peak positions and the normal-state
dispersion.
\begin{figure}[t]
\begin{center}
\includegraphics[width=1\textwidth]{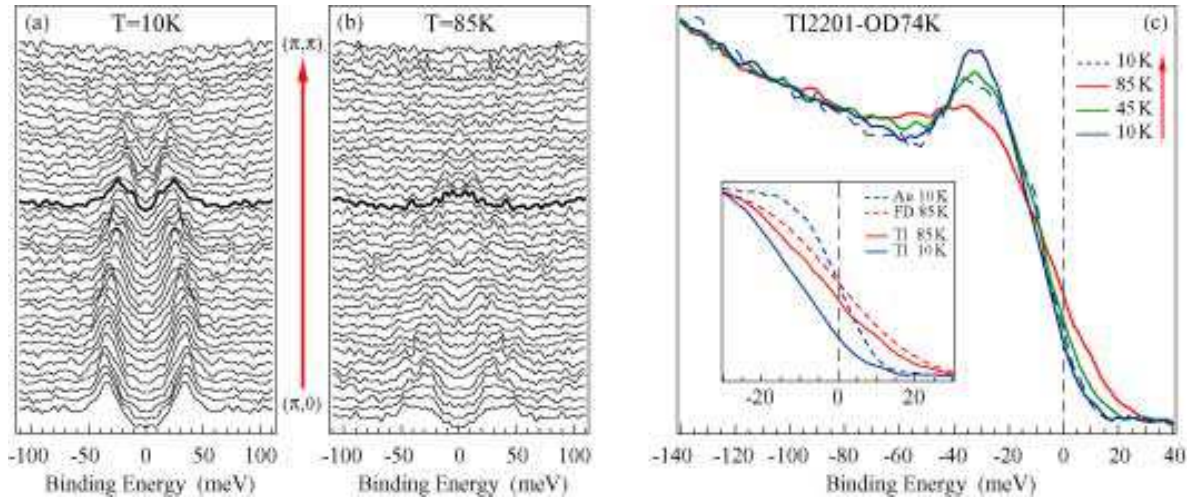}
\caption{\label{Tl-temp-gap-symm74} Symmetrized ARPES spectra measured at (a) 10\,K and (b) 85\,K on overdoped Tl2201-OD74, along the
$(\pi,0)\!-\!(\pi,\pi)$ direction. Bold lines indicate the EDCs closest to $k_F$. (c) The $k$ integrated spectral weight at the antinode, for
several temperatures, together with a 10\,K Au Fermi edge and an 85\,K resolution-broadened Fermi function; an enlarged view of the leading edge
positions is given in the inset.}
\end{center}
\end{figure}
At lower binding energies, however, the peak does not reach $E_F$, but instead reaches a minimum at $\Delta_P\!\simeq\!17$\,meV and then
disperses back to higher binding energy. This behavior is a hallmark of Bogoliubov QPs in a superconductor, as shown in the sketch of
Fig.\,\ref{Tl-gap-fitted}(f,g), for which a beautiful experimental demonstration was recently obtained by ARPES on the trilayer Bi-cuprate
Bi$_2$Sr$_2$Ca$_2$Cu$_3$O$_{10+\delta}$ \cite{Matsui:2003}. Simultaneously, we observe a reduction of spectral weight, by about a factor of 2,
when the QP peak has dispersed from the van Hove singularity all the way to $k_f$. This is consistent with the spectral intensity of Bogoliubov
QPs being determined by the coherence factor $v_{\bf k}^2\!=\!1\!-\!u_{\bf k}^2\!=\!\frac{1}{2}(1\!-\!\epsilon_{\bf k}/E_{\bf k})$, with $E_{\bf
k}\!=\!\sqrt{\epsilon_k^2\!+\!\Delta_{\bf k}^2}$, which corresponds to $1/2$ when $\epsilon_{\bf k}\!=\!\epsilon_{{\bf k}_f}\!=\!0$ and 1 for
$|\epsilon_{\bf k}|\!\gg\!\Delta_{\bf k}$.  Extending this analysis to other momenta along the Fermi surface, one can study the $k$-dependence
of the gap: at $\sim\!(\pi/2,\pi/2)$ the peak does cross $E_F$ (not shown), while at intermediate momenta a gap smaller than at $(\pi,0)$ is
observed, as on the right-hand side of Fig.\,\ref{Tl-gap-fitted}(d), consistent with $d$-wave symmetry.

The temperature dependence of the gap at the antinode has been measured for Tl2201-OD74. Fig.\,\ref{Tl-temp-gap-symm74}(a,b) present symmetrized
ARPES data along the $(\pi,0)\!-\!(\pi,\pi)$ direction, at temperatures above and below $T_c$. In exactly the same manner as the $k$-dependent
symmetrized data of Fig.\,\ref{Tl-gap-symmetrized}(a,b), the temperature-dependent data of Fig.\,\ref{Tl-temp-gap-symm74}(a,b) show a clear gap
at $k\!=\!k_F$ at 10\,K, which is clearly smaller at 85\,K (with this type of analysis, the broadness of the QP peak and noise level at 85\,K
make it hard to conclude whether the gap has completely closed).
\begin{figure}[t]
\begin{center}
\includegraphics[width=1\textwidth]{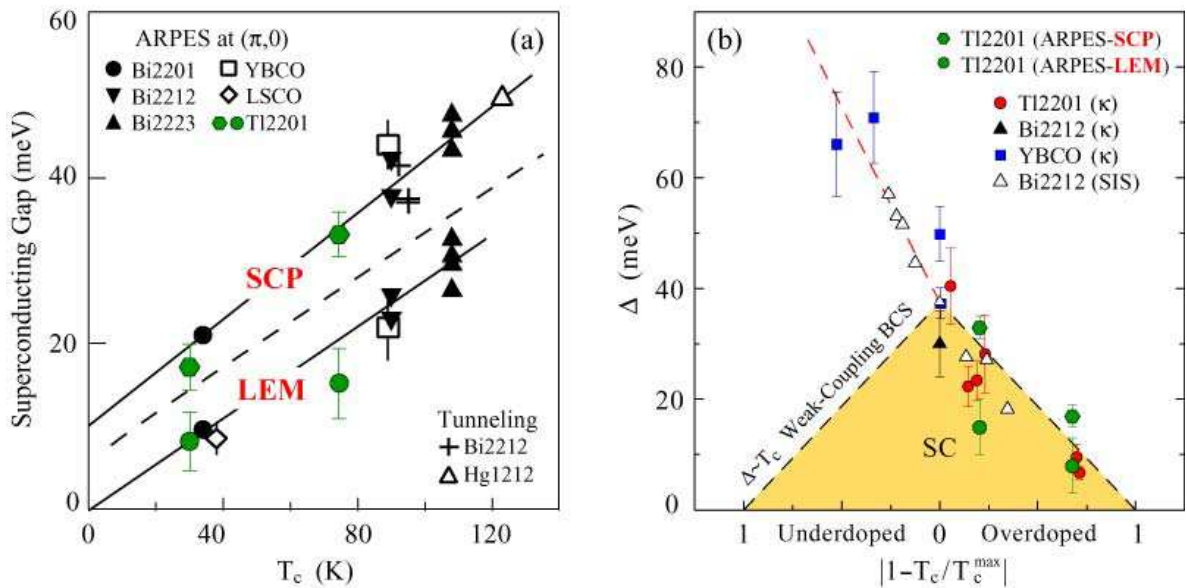}
\caption{\label{gap-compilation} (a) Superconducting gap magnitude estimated by tunneling and ARPES from the superconducting peak position (SCP)
and the leading-edge midpoint (LEM) shift, plotted versus $T_c$ for various optimally or overdoped cuprates (after Ref.\,\cite{Feng:2002}). (b)
Gap values from thermal conductivity, tunneling, and ARPES plotted vs.\ $|1\!-\!T_c/T_c^{max}|$ for many HTSCs across the phase diagram (after
Ref.\,\cite{Hawthorn}); dashed lines are guides to the eye. Both plots include our own Tl2201-OD74 and Tl2201-OD30 ARPES data.}
\end{center}
\end{figure}
Fig.\,\ref{Tl-temp-gap-symm74}(c) shows averaged EDCs taken over a narrow $k$-range around $k_f$ at 10, 45, 85, and again at 10\,K. Included in
the figure is a 10\,K Au reference spectrum, and the simulation of an energy-resolution broadened Fermi function at 85\,K. The inset of
Fig.\,\ref{Tl-temp-gap-symm74}(c) shows a closer view of these $k$-integrated EDCs around E$_f$. It is clear from this figure that, as the
temperature increases, the size of the gap decreases although it does not seem to completely close at 85\,K. This is suggested by comparing the
position of the leading edge of the 85\,K ARPES data with the simulated Fermi function; they are close in energy but not quite coincident. It is
possible that there is still a pseudogap at temperatures slightly above $T_c\!=\!74$\,K at this doping level. It is worth emphasizing that as
the temperature is lowered back to 10\,K, the gap reopens to approximately the same size. Although a partial degradation of the ARPES features
can be seen, the QP peaks and the gap are still clearly observable. This is strong evidence that the surface is stable under these experimental
conditions and is thus suitable for detailed surface-sensitive experiments. From these data, a superconducting peak (SCP) position of
$\sim\!33$\,meV and a leading-edge midpoint (LEM) gap of $\sim\!15$\,meV can be extracted.

As a direct comparison with other cuprates and different means of determining the gap, Fig.\,\ref{gap-compilation}(a,b) present a compilation of
the doping dependence of the gap magnitude. Fig.\,\ref{gap-compilation}(a) refers only to spectroscopic studies of optimally and overdoped
HTSCs, for which the gap is proportional to $T_c$. Our results from Tl2201-OD74 and Tl2201-OD30 are fully consistent with the behavior observed
on the other cuprates, as far as both SCP and LEM positions are concerned. Even more significant is Fig.\,\ref{gap-compilation}(b), which
demonstrates that ARPES measures gap values that follow the same trend as those derived from thermal conductivity, a bulk transport property. A
more accurate comparison would require an analysis of the ARPES data beyond the mere determination of SCP and LEM positions, for instance on the
basis of a model spectral function; this is however complicated by the non-trivial doping dependence of the ARPES data, and in the present case
was done for Tl2201-OD30 but not Tl2201-OD74.

\subsection{ARPES Lineshape Analysis}
\label{lineshape}

The electrical and thermal transport properties suggest that overdoped Tl2201 may be regarded as a Fermi liquid
\cite{Mackenzie:1996,Proust:2002,Hussey:2003}. Since the ARPES data for both the normal and superconducting states indicate that single-particle
surface-sensitive probes do provide information representative of the bulk, overdoped Tl2201 might be an ideal material for a quantitative study
of the strength and nature of the many-body effects, which are revealed by the energy and momentum evolution of a QP's intrinsic ARPES lineshape
\cite{AndreaScripta}. In general, however, the shape of an ARPES peak does not correspond directly to an excitation's intrinsic lineshape. This
is primarily due to an extrinsic contribution from momentum and energy resolution of the analyzer. Recent work on the model two-dimensional
Fermi liquid system Sr$_2$RuO$_4$ highlights the significant issues involved in removing the effects of the analyzer resolution \cite{Ingle}; in
particular, it clearly shows that both the lineshape width and the peak position can be affected by resolution effects. There are significant
challenges to completely removing the analyzer resolution from the data and stringent requirements on the data quality needed to attempt such
analysis; therefore, we will analyze the current Tl2201 spectra in a more qualitative and phenomenological manner, concentrating primarily on
the momentum dependence of the QP peaks measured in the superconducting state for different doping levels.

Beyond the issue of the experimental resolution, matrix element effects and the handling of the ARPES background are the two main complications
in attempting even a qualitative description of the momentum dependence of the QP lineshapes.  Experimentally, problems due to matrix element
effects may be reduced by appropriate choice of measurement conditions such as geometry, photon energy, and polarization. For the experiments
performed at the SLS and the data discussed in the following, circularly polarized 59\,eV photons were used. Circular polarization was chosen
specifically to minimize matrix element effects, which, using symmetry arguments, can be shown to be most extreme for linear polarization
\cite{Andrea:2003}. The photon energy was selected based on our experience on other cuprates
--- specifically, La$_{2-x}$Sr$_x$CuO$_4$ (LSCO) where similar photon energies were used to
reveal structure that had previously been missed in lower-energy ARPES data
\cite{Lureview,Zhoureview}. Finally, in order to gain more reliable information, the ARPES
data were taken over multiple Brillouin zones; this way it was possible to verify that, for
the same point in the reduced zone scheme, the lineshapes from different zones in the
extended zone scheme were qualitatively the same.

The most commonly used method to isolate a QP peak from the experimental background \cite{Andrea:2003,Kaminski:2004}, is to subtract a
phenomenological ARPES background determined from the data at $k$-values far removed from those at which the QP-like peaks are detected.
\begin{figure}
\begin{center}
\includegraphics[width=5in]{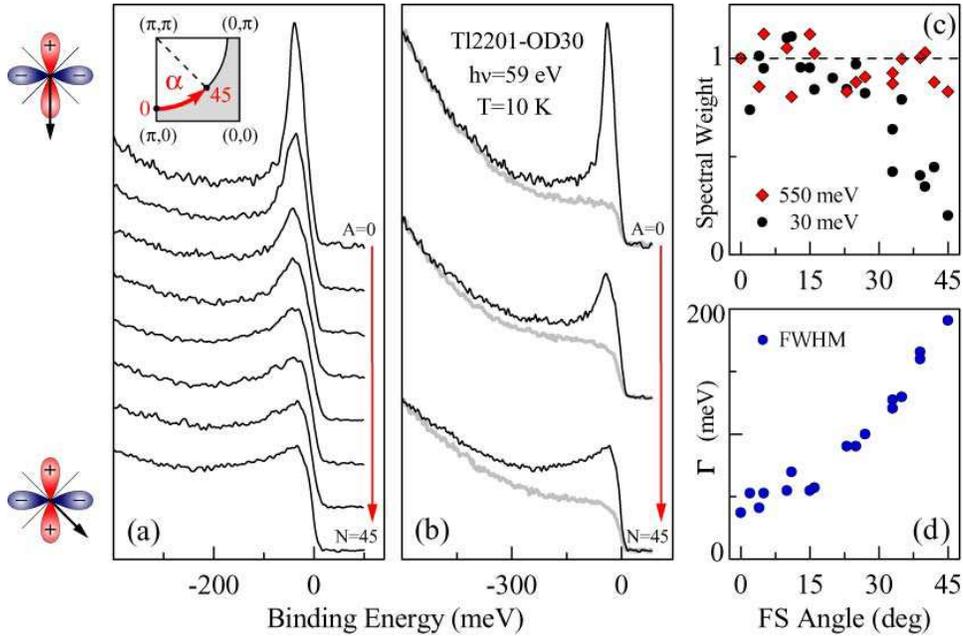}
\caption{(a) Tl2201-OD30 ARPES spectra at $k$ slightly smaller than $k_F$ along the Fermi surface contour (corresponding to a QP binding energy
of $\sim$35\,meV). (b) Selected spectra from (a) along with their corresponding $k\gg k_F$ background. (c) Spectral weight of the background
subtracted spectra, integrated over different energy ranges and normalized with respect to the $\alpha\!=\!0$ antinodal value, plotted vs.\ the
Fermi-surface angle $\alpha$. (d) Quasiparticle linewidth $\Gamma$ plotted vs.\ the Fermi-surface angle $\alpha$. \label{fig:Andrea3}}
\end{center}
\end{figure}
In our analysis we will use this approach, taking as background the weakly $k$-dependent photoemission intensity observed in each momentum space
cut for $k\gg k_F$. The fundamental problem with this approach is that if some of the background is related to the incoherent part of the
spectral function, as one would expect for correlated electron system, then by subtracting the background we are actually disregarding some of
the most crucial information contained in the ARPES spectra. In fact, the $k$- and $\omega$-dependence of the incoherent part of the spectral
function is, in many respects, as important as the behavior of the QP peak itself; unfortunately we do not yet have the means to reliably
extract information from it.

In Fig.\,\ref{Tl-QP}(b), similar to what was previously observed in all photoemission studies of the cuprates \cite{Andrea:2003,Campuzano}, we
saw that in the nodal region the width of the QP peaks increases as a function of binding energy, as expected from simple phase-space arguments.
This is, however, in sharp contrast to what is observed in the antinodal region, where the sharpest peak is found at the bottom of the band: in
Fig.\,\ref{Tl-gap-fitted}(c), the linewidth $\Gamma$ of the QP peak is observed to grow from $\sim$30 to $\sim$55\,meV as the QP peak disperses
from $\sim$39 to $\sim$20\,meV binding energy. Possible origins for this anomalous behavior will be discussed later, but it should be pointed
out that an analogous effect was recently reported for very overdoped and nearly non-superconducting
Bi$_{1.74}$Pb$_{0.38}$Sr$_{1.88}$CuO$_{6+\delta}$ \cite{donglai:2006}. Another striking feature of the lineshape evolution is that the QP peaks
are much broader in the \emph{nodal} region than in the \emph{antinodal} region, as evidenced by the direct comparison of Fig.\,\ref{Tl-QP}(b)
and\,\ref{Tl-QP}(c).
\begin{figure}
\begin{center}
\includegraphics[width=1\textwidth]{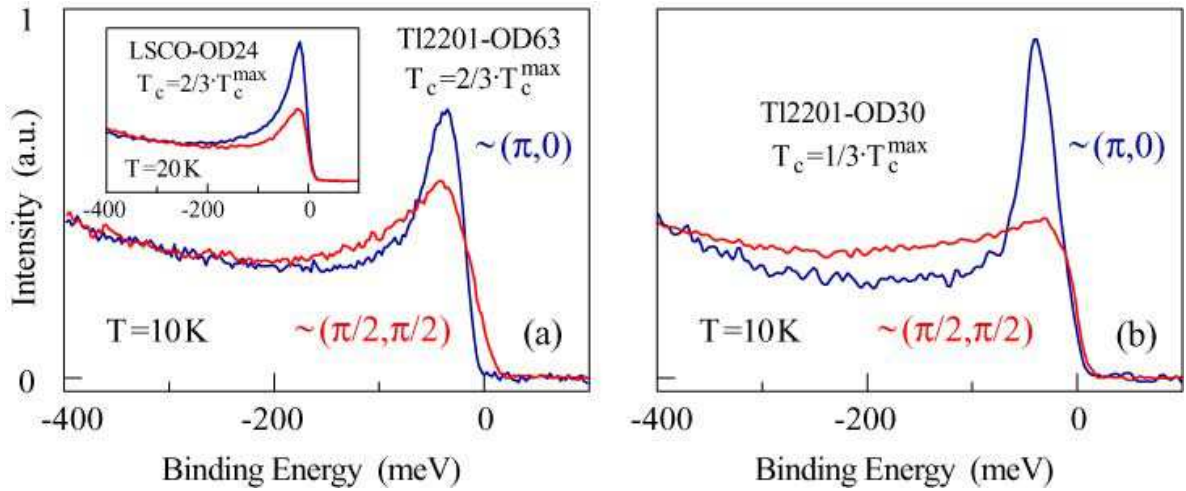}
\caption{\label{fig:Andrea4} Nodal and antinodal spectra from (a) Tl2201-OD63 and (b) Tl2201-OD30, at $k\!\lesssim\!k_F$. Inset: similar data
from overdoped LSCO-OD24 ($p\!=\!0.22$), from Ref.\,\cite{Zhou:2004}. For both samples in (a), $T_c\!\simeq\!\frac{2}{3}T_c^{max}$ with respect
to the $T_c^{max}$ of each family \cite{Eisaki:2004}.}
\end{center}
\end{figure}
To elaborate on this, in Fig.\,\ref{fig:Andrea3}(a) we present a compilation of spectra from along the Fermi surface contour, with $k$ slightly
smaller than $k_F$ and corresponding to a binding energy of $\sim$35\,meV. The choice in favor of spectral peaks at a binding energy lower than
$E_F$ is dictated by the need to compare spectra not affected by either the opening of the $d$-wave gap or the anomalous low-binding-energy
broadening seen at the antinode. The sharp peak near $(\pi,0)$ becomes progressively broader as $(\pi/2,\pi/2)$ is approached. In order to
determine whether this apparent broadening represents an increase in the QP linewidth $\Gamma$, or is merely a loss of spectral weight, the
momentum-independent background from $k\gg k_F$ is subtracted from the ARPES spectra as shown in Fig.\,\ref{fig:Andrea3}(b); then the data are
integrated over varying energy ranges and plotted in Fig.\,\ref{fig:Andrea3}(c), normalized at the Fermi surface angle $\alpha\!=\!0^{\circ}$.
For narrow integration windows, there is a drop as a function of $\alpha$: the QPs show a loss of low-energy weight (and possibly coherence) in
the nodal region. However, once the integration window is expanded to about 550\,meV, the spectral weight of the QP peaks becomes
angle-independent. This indicates that the $k$-dependent broadening of the ARPES spectra indeed reflects a loss of coherence of the QP spectrum,
rather than being due to matrix element effects.

The observed QP anisotropy of Tl2201-OD30, with peaks much broader in the nodal than the antinodal region, is in sharp contrast to the behavior
observed in underdoped cuprates, where the QP peaks are sharp near $(\pi/2,\pi/2)$ and very broad around $(\pi,0)$, as summarized in
Fig.\,\ref{anisotropy}.
\begin{figure}
\begin{center}
\includegraphics[width=1\textwidth]{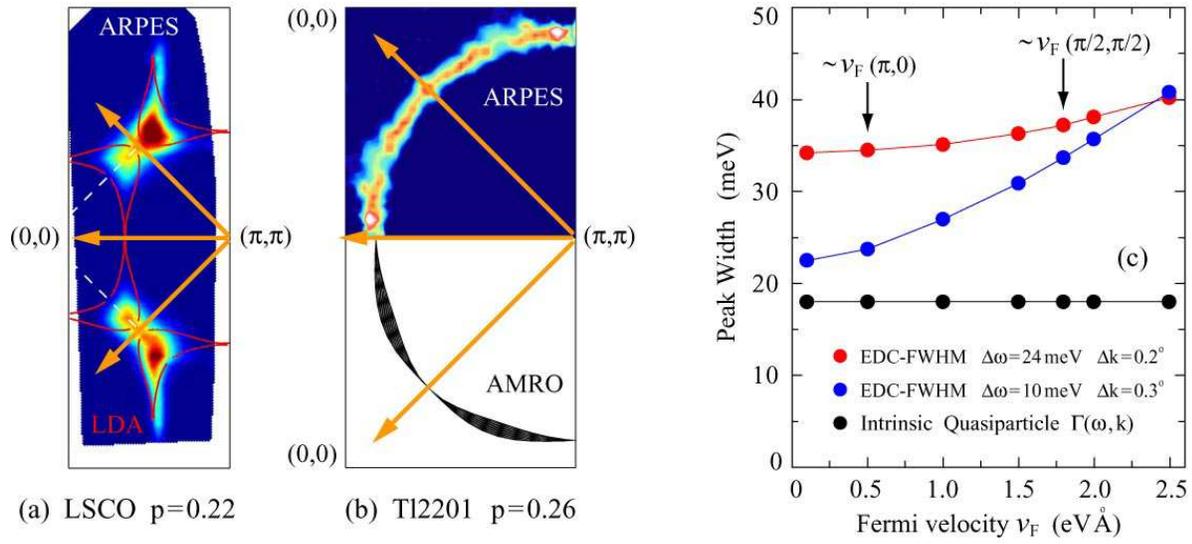}
\caption{\label{fig:Tl2201_NJP_broad} (a) ARPES intensity map for emission from the Fermi energy for overdoped LSCO-OD24 ($p\!=\!0.22$). Red
lines are the calculated LDA three-dimensional Fermi surface for $k_z\!=\!0$ and $\pi /c$; the enclosed area is the projection of the Fermi
surface onto the two-dimensional $k_x$-$k_y$ plane, and denotes the region allowed for emission \cite{SahrakorpiLSCO}. (b) ARPES and AMRO data
from Tl2201-OD30. In (a,b) the yellow arrows identify directions characterized by zero $k_z$ dispersion as indicated by the LDA calculations and
consistent with AMRO experiments, which also suggest much weaker residual three-dimensionality effects in Tl2201 than in LSCO. (c) Resolution
contributions to the width of an ARPES EDC calculated, as a function of the Fermi velocity $v_F$, from the two-dimensional convolution of a
Lorentzian quasiparticle peak (binding energy $\omega\!=\!30$\,meV and intrinsic width $\Gamma\!=\!18$\,meV) with Gaussian energy and momentum
resolution functions (results corresponding to our experimental parameters are in red).}
\end{center}
\end{figure}
As doping is increased, the antinodal QP peaks sharpen but remain significantly broader than the nodal QPs up to optimal doping, at least in the
normal state. Even in the superconducting state, where antinodal QPs sharpen considerably, the ARPES linewidths for underdoped and
optimally-doped materials are still highly anisotropic, with a minimum at $(\pi/2,\pi/2)$ \cite{Andrea:2003,Campuzano}. Overall this behavior
leads to the expectation that increasing doping beyond the optimal value will simply render the QP linewidths progressively more isotropic.
Indeed, isotropic lineshapes were observed on early data from very overdoped, non-superconducting
Bi$_{1.80}$Pb$_{0.38}$Sr$_{2.01}$CuO$_{6-\delta}$ \cite{Kaminski:2005}, although more recent work on the same material \cite{donglai:2006}
exhibits a behavior more consistent with what is reported here for Tl2201. Going back to the QP anisotropy reversal observed for Tl2201-OD30 in
Fig.\,\ref{fig:Andrea3}, and presented again in Fig.\,\ref{fig:Andrea4}(b) for nodal and antinodal points only, it must be emphasized that the
same, albeit less pronounced, effect was observed in our Tl2201-OD63 sample, as shown in Fig.\,\ref{fig:Andrea4}(a). For comparison, in the
inset of Fig.\,\ref{fig:Andrea4}(a) we have plotted ARPES data from overdoped ($x\!=\!p\!=\!0.22$) La$_{2-x}$Sr$_x$CuO$_{4}$ \cite{Zhou:2004}.
This sample has $T_c\!\simeq\!\frac{2}{3} T^{{max}}_c$, comparable to the degree of overdoping of our sample Tl2201-OD63, which also has
$T_c\!\simeq\!\frac{2}{3} T^{{max}}_c$ for the much larger $T_c^{max}$ of the Tl2201 family. The data from LSCO exhibit a QP anisotropy similar
to that in Tl2201-OD63, suggesting that the observed QP anisotropy reversal might indeed be generic to the overdoped cuprates.

Before proceeding to the discussion of the broader significance of these findings, we comment here on two extrinsic effects that could
potentially contribute to an anomalous $k$-dependent broadening of the QP lineshapes: residual $k_z$ electronic dispersion and resolution
broadening effects. First, the dispersion of the electronic structure along the $c$-axis may indeed give rise to $k_{||}$-dependent broadening
of the ARPES features \cite{SahrakorpiLSCO} (where $k_{||}$ is the $ab$-plane momentum). This effect could be doping-dependent because there is
a 0.4\% $c$-axis lattice constant reduction as doping is increased \cite{Wagner}. However, the $k_z$-dispersion vanishes in Tl2201 along the
$(0,0)\!-\!(\pi,\pi)$ and $(\pi,0)\!-\!(\pi,\pi)$ directions, and thus at all nodal and antinodal points. This is a direct consequence of the
symmetry of the body-centered tetragonal unit cell, and holds also for LSCO. Fig.\,\ref{fig:Tl2201_NJP_broad}(a,b) illustrate this point by
showing ARPES data for both materials together with band structure calculations for LSCO \cite{SahrakorpiLSCO}, and AMRO data for Tl2201
\cite{Hussey:2003}. Along these high-symmetry directions, no $k_z$-dispersion is present either in the LSCO calculations or the Tl2201 AMRO data
(in both cases the $k_z$-dispersion is proportional to the width of the three to two-dimension projected contours). This is consistent with the
ARPES maps for the two systems, although the broader $E_F$ intensity patches for LSCO indicate that the three-dimensionality is much stronger in
LSCO than in Tl2201. A residual $k_z$ dispersion might contribute to the overall width of the ARPES features, but it cannot be responsible for
the reversal of the anisotropy observed on overdoped cuprates, since the anisotropy exhibits a monotonic $k_{||}$-dependence between
$(\pi/2,\pi/2)$ and $\sim\!(\pi,0)$.

The second extrinsic effect that influences the width in energy of an ARPES spectrum, or EDC, is instrumental resolution. In a two-dimensional
system, the width in energy is determined by the intrinsic inverse lifetime of the QP excitation and by the experimental resolution broadening,
which reflects both the energy ($\Delta\omega$) and angular ($\Delta k$) resolution of the apparatus \cite{AndreaScripta}. For a dispersionless
feature, the instrumental contribution to the total EDC width is determined solely by the energy resolution $\Delta\omega$. However, for a
dispersive feature, the angular resolution $\Delta k$ contributes to the energy broadening in a manner directly proportional to the band
velocity ($v_F\!\neq\!0$ in Fig.\,\ref{fig:Tl2201_NJP_broad}(c)). Since the Fermi velocity around the Fermi surface of the cuprates is highly
anisotropic, varying from approximately 0.5 to 1.8\,eV\AA\ in going from antinodal to nodal region, this should give rise to a
momentum-dependent resolution broadening of the EDCs. This broadening would be more severe at the nodes, where $v_F$ is at its largest, which
raises the concern that the larger nodal widths shown in Fig.\,\ref{fig:Andrea3} and\,\ref{fig:Andrea4} might be an artifact of the instrumental
resolution. We have carefully evaluated the contribution of resolution broadening effects for our experimental conditions with a variety of
different procedures and the results are summarized in Fig.\,\ref{fig:Tl2201_NJP_broad}(b). For our experimental parameters (red data,
$\Delta\omega\!=\!24$\,meV and $\Delta k\!=\!0.3^{\circ}$) the $k$-dependent resolution broadening is of the order of few meV, nowhere near the
factor of 4 observed in Fig.\,\ref{fig:Andrea3}(d).

\section{Discussion and Conclusions}
\label{discussion}

Using a self-flux method and partial encapsulation, we have grown single crystals of the single-layer overdoped cuprate superconductor Tl2201.
The crystals were annealed under controlled oxygen partial pressures to set a homogeneous doping level whilst preventing decomposition, and
typically exhibited sub-Kelvin superconducting transition widths as measured in a SQUID magnetometer. Their high quality and homogeneity were
evidenced not only by these narrow transition widths but also by their extremely narrow rocking curve widths, comparable to
YBa$_2$Cu$_3$O$_{7-\delta}$ grown in YSZ crucibles \cite{Ruixing:1998}. An EPMA study found the crystals to have the composition
Tl$_{1.884(6)}$Ba$_2$Cu$_{1.11(1)}$O$_{6+\delta}$, which was homogeneous and was not observed to vary between crystals. This level of cation
substitution is similar to that reported earlier for single crystals of Tl2201. For higher dopings (lower $T_c$s), the crystals were found to be
orthorhombic, a slight deviation from a square-planar lattice that is often associated with the unsubstituted phase which has only been prepared
in ceramics. The orthorhombic distortion appears to be suppressed by cation disorder, but stabilized by the oxygen interstitials that do
increase the doping into the very overdoped regime.

While these crystals constitute a significant step forward, much optimization of the growth technique remains, and there is still the challenge
of growing single crystals without the cation disorder in the Tl layer. Still, the high quality Tl2201 single crystals grown as part of this
effort have enabled the first successful ARPES experiments of this compound. The results on the Fermi surface of Tl2201 mark an important new
starting point for understanding the cuprates, namely that there is a material where both a surface-sensitive single-particle spectroscopic
technique (ARPES) and comparable bulk transport measurements (AMRO) have arrived at quantitative agreement on a major feature of the normal
state \cite{Hussey:2003,mauro}.
\begin{figure}
\begin{center}
\includegraphics[width=1\textwidth]{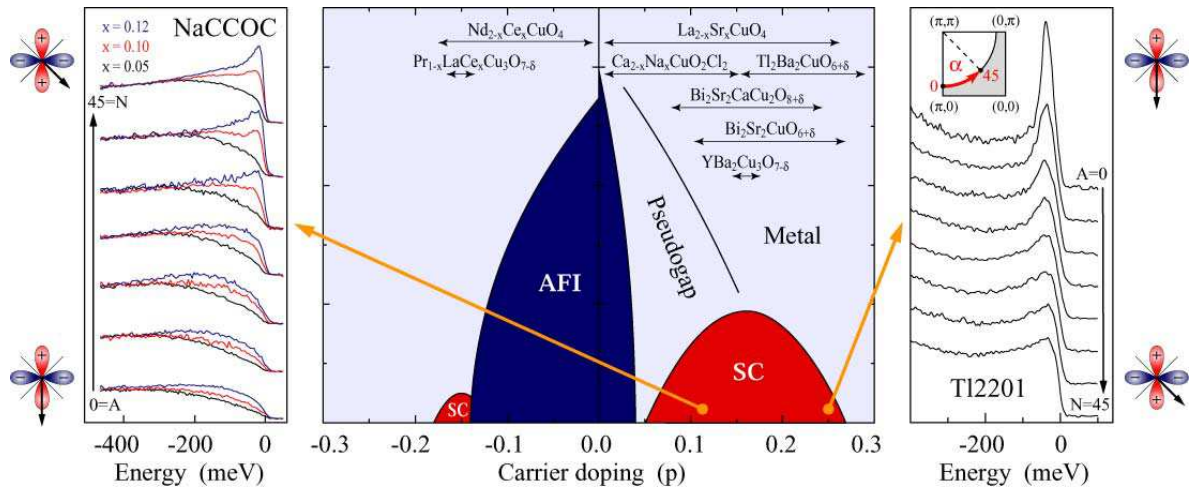}
\caption{\label{anisotropy} Nodal-antinodal quasiparticle anisotropy reversal observed in the cuprates. As a function of the Fermi surface angle
$\alpha$, sharp QP peaks are observed at the nodes (N) and ill-defined QPs at the antinodes (A) in underdoped cuprates (e.g., data from
Ca$_{2-x}$Na$_x$CuO$_2$Cl$_2$, after Ref.\,\cite{2}); in overdoped Tl2201, this behavior is reversed.}
\end{center}
\end{figure}
This is a first for the copper-oxide high-$T_c$ superconductors and, within the more general class of 3$d$ and 4$d$ transition metal oxides, it
is second only to the case of Sr$_2$RuO$_4$ for which a similar quantitative agreement was reached between de Haas-van Alphen
\cite{Mackenzie:sro,bergemann:sro} and ARPES \cite{Ingle,Damascelli:2000,Liebsch:comm,Damascelli:comm,Shen:sro}. The detailed agreement on the
Fermi surface (Fig.\,\ref{FS}), together with the good quantitative agreement achieved for the superconducting gap by thermal conductivity and
ARPES (Fig.\,\ref{gap-compilation}), establishes Tl2201 as an ideal system to study the overdoped regime of the cuprate phase diagram with a
wide spectrum of techniques. Next it will be important to study the single-particle excitation spectrum with both ARPES and STS in enough detail
to compare to other normal state properties, thus critically testing whether or not a conventional Fermi-liquid description of the electronic
properties captures the physics of the normal metal in the heavily overdoped regime. In this regard, it is important to elaborate on the most
surprising result that emerged from this study of overdoped Tl2201 by ARPES, one which is counterintuitive within the realm of Fermi liquid
theory: the quasiparticle anisotropy reversal observed across optimal doping. A summary of the momentum anisotropy in under and overdoped
cuprates is given in Fig.\,\ref{anisotropy}, where data from Ca$_{2-x}$Na$_x$CuO$_2$Cl$_2$ \cite{2}, for several doping levels, and Tl2201 are
directly compared.

Early magnetotransport experiments on very overdoped Tl2201, for which a small magnetoresistance and a weak $T$-dependence of the resistivity
and cotangent of the Hall angle have been observed, do not support the presence of strongly $k$-dependent low-temperature scattering rates in
the normal state \cite{Mackenzie:1996}. More recently, a temperature, doping, and magnetic field dependent AMRO study of Tl2201 suggested a
two-component scattering rate consisting of an isotropic $T^2$ term as well as an anisotropic $T$-linear term \cite{Nigelp}, which vanishes at
the nodes and has a maximum near the antinodes. The anisotropic term becomes weaker upon overdoping the material, which is consistent with the
view that as hole doping increases, the antinodal QPs should gain coherence faster than the nodal ones, eventually leading to a relatively small
and fully isotropic scattering rate. Therefore, what is at variance between the ARPES and AMRO results from Tl2201 is not the behavior of the
antinodal QPs, which in both cases are indeed gaining coherence upon overdoping, but rather that of nodal QPs which in ARPES become
progressively less coherent. In addition, the overall magnitude of the scattering rate seen by ARPES is approximately a factor of 10 larger than
estimated by AMRO.

What could be the origin for this discrepancy between ARPES and AMRO determined scattering rates? First of all, one has to note that while the
ARPES experiments are performed in the superconducting state, the AMRO measurements are carried out in the normal state; and the latter is
achieved by the application of external magnetic fields as high as 45T, which might potentially affect the low-energy QP dynamics. As for the
reliability of the ARPES data on this specific point of the loss of coherence of nodal QPs at large dopings, it is important to mention that
these findings are also supported by a preliminary study of overdoped Bi$_2$Sr$_2$CaCu$_2$O$_{8+\delta}$ by STM and STS \cite{Jamesp}. From the
Fourier analysis of the energy-dependent spatial modulations observed in the tunneling conductance \cite{Hoffman}, it was concluded that for
strongly overdoped samples, the nodal (i.e., low energy) quasiparticle interference signal is no longer visible, consistent with a decoherence
of the nodal states \cite{Jamesp}.

A reconciliation of the results might require consideration of the specific sensitivity of transport probes and single-particle spectroscopies
to electronic scattering phenomena. In particular, photoemission and scanning tunneling are very sensitive to electronic scattering involving
small momentum transfer, while transport is rather insensitive to it. A possible source of such scattering could be extended impurities far from
the CuO$_2$ planes \cite{varma}, such as cation substitution or interstitial oxygen which increase with doping in most cuprates. Recent
calculations show that the inclusion of elastic forward scattering in a $d$-wave superconductor can lead to an overall increase of the
scattering rate in the normal state, but also to a strong enhancement of antinodal QP coherence in the superconducting state \cite{Zhu:2004}. A
very different model, based on the description of unitary-limit scattering beyond the Born approximation so as to account for multiple
scattering from a single impurity, would also lead, below $T_c$, to loss of nodal coherence and anomalous enhancement of antinodal coherence
\cite{wakabayashi}. However, both models predict an approximately isotropic QP scattering above $T_c$, as well as temperature dependent effects
limited to the energy scale of the gap itself. To test the applicability of these scenarios to the present case, a thorough temperature
dependent study is required. Finally, it should also be pointed out that sources of scattering relevant to this discussion are not just limited
to impurities; also inelastic (i.e., electronic) scattering involving small momentum transfer might result in anomalous decoherence of the nodal
QPs. For instance, low-energy (i.e., small $q$) quantum-critical fluctuations associated with proximity to a competing superconducting
$d_{x^2-y^2}\!+\!id_{xy}$ phase \cite{Vojta:2000}, or ferromagnetic phases, could also lead to a similar QP anisotropy reversal
\cite{Chakravarty}.

\ack We gratefully acknowledge D.G. Hawthorn, K.M.\ Shen, N.E.\ Hussey, A.P.\ Mackenzie, J.C. Davis, D.J.\ Scalapino, and G.A.\ Sawatzky for
discussions and M.\ Plat$\acute{{\rm e}}$, N.P.\ Armitage, A.B.\ Kuzmenko, S.\ Chiuzbaian, M.\ Shi, M.\ Falub, and L.\ Patthey for assistance
during the ARPES experiments. We are also grateful to machine and beamline groups at the Swiss Light Source and Stanford Synchrotron Radiation
Laboratory, whose outstanding efforts made the experiments possible. This work was supported by the Canada Research Chairs Program, the Natural
Sciences and Engineering Research Council of Canada, the Canadian Institute for Advanced Research, and the British Columbia Synchrotron
Institute.

\end{document}